\documentclass[aps,pra,pdflatex,reprint,10pt,showpacs,superscriptaddress,floatfix]{revtex4-2}
\usepackage{float}
\usepackage{amsmath}
\usepackage[utf8x]{inputenc}
\usepackage{natbib}
\usepackage{graphicx}
\usepackage{hyperref}
\usepackage{epstopdf}
\usepackage{pifont}
\usepackage{float}
\usepackage{ulem}
\usepackage{placeins}

\graphicspath{{pdf/}}
\hypersetup{colorlinks=true,linkcolor=blue,citecolor=blue,filecolor=blue,urlcolor=blue,breaklinks=true}
\RequirePackage{color}

\usepackage{float}

\begin{document}

\date{\today}

\title{Exact treatment of rotation-induced modifications in two-dimensional quantum rings}

\author{Carlos Magno O. Pereira}
\email[Carlos Magno O. Pereira - ]{carlos.mop@discente.ufma.br}
\affiliation{Departamento de F\'{\i}sica, Universidade Federal do Maranh\~{a}o, 65085-580 S\~{a}o Lu\'{\i}s, Maranh\~{a}o, Brazil}

\author{Frankbelson dos S. Azevedo}
\email[{Frankbelson dos S. Azevedo - }]{frfisico@gmail.com}
\altaffiliation{On leaving the current affiliation and currently unaffiliated.}
\affiliation{Departamento de F\'{\i}sica, Universidade Federal do Maranh\~{a}o, 65085-580 S\~{a}o Lu\'{\i}s, Maranh\~{a}o, Brazil}

\author{Edilberto O. Silva}
\email[Edilberto O. Silva - ]{edilberto.silva@ufma.br}
\affiliation{Departamento de F\'{\i}sica, Universidade Federal do Maranh\~{a}o, 65085-580 S\~{a}o Lu\'{\i}s, Maranh\~{a}o, Brazil}

\begin{abstract}
We investigate the influence of rotation on the Fermi energy, magnetization, and persistent current in two-dimensional quantum rings. Using the Tan–Inkson confinement potential and incorporating rotational effects through a non-inertial coupling, we derive analytical expressions for the energy levels and examine the modifications induced by rotation. We then numerically explore how variations in angular velocity affect the Fermi energy, magnetization, and persistent current. Our results show that rotation has a significant impact on these physical properties, underscoring the importance of considering rotational effects in quantum ring systems. This suggests that rotation could serve as a control parameter in the development of new mesoscopic devices, without the need for additional fields or geometric modifications.
\end{abstract}

\maketitle

\section{Introduction}

Quantum Rings (QRs) are nanostructures with a unique topology that gives rise to remarkable electronic, magnetic, and optical properties. These properties are particularly pronounced under the influence of an external magnetic field, which induces an Aharonov–Bohm (AB) flux through the ring \cite{SST.2002.17.22,PR.1995.115.485}. Both the magnetic field and the resulting AB flux significantly impact quantum interference effects in these systems \cite{PR.1995.115.485,W-CTan_1996}. Recent advances in growth techniques have enabled the fabrication of self-organized QRs with diameters on the order of tens of nanometers \cite{N.2023.12.1497,PRL.2000.84.2223,N.2001.413.825}. Their distinctive characteristics make them promising candidates for technological applications and ideal platforms for studying fundamental quantum-mechanical phenomena~\cite{fomin2013physics}. In particular, QR systems have demonstrated the ability to control electronic properties through precise adjustments, such as tuning the average ring size or applying external magnetic fields \cite{HERNANDEZ2025116122,NAIFAR2025208145,Shirsefat_2025,SHAER2025417145,PhysRevB.111.125409,BAKDID2025172891,AYALAQUITIAQUEZ2024129980,10.1063/10.0028635}. These manipulations lead to intriguing phenomena and offer exciting opportunities for tailoring the quantum behavior of QRs~\cite{nano13091461,10.1063/10.0020602}.

Quantum mechanics in rotating frames reveals phenomena that are absent in classical systems. Rotation introduces additional terms in Maxwell’s equations \cite{WM.2017.72.69,PRB.2024.110.035129} and the Dirac equation \cite{U.2024.10.389}, with dependencies on angular velocity. These effects are relevant to Earth-based experiments, magnetic resonance physics, and superconducting systems~\cite{https://doi.org/10.1002/andp.19734840109}. Quantum interference experiments have demonstrated phase shifts resulting from gravitational and rotational effects. These shifts correspond to classical transverse accelerations in rotating frames and reflect the equivalence principle as applied to quantum systems~\cite{PhysRevD.15.1448}. The Sagnac phase shift and spin-rotation coupling can be derived in the non-relativistic limit, while a novel phase shift due to spin–orbit coupling emerges from the low-energy limit of the Dirac equation with spin connection~\cite{Anandan2004}.

In the context of QR semiconductors, recent studies have examined the effects of rotation, revealing significant impacts on their physical properties. Rotation notably influences thermodynamic quantities such as mean energy, specific heat, and entropy~\cite{Ghanbari2024RotatingEO}. It also modifies the probability distribution of charged particles in two-dimensional rings, causing electrons to be pushed toward the edges and altering optical properties, including the refractive index and absorption coefficients~\cite{Lima2023OpticalAE}. Furthermore, rotation lifts the degeneracy of Landau levels in 2D rings and induces Aharonov–Bohm oscillations in magnetization, which are associated with edge states~\cite{https://doi.org/10.1002/andp.202200371}. The photoionization process in such quantum systems has also been shown to depend on parameters such as the average radius, AB flux, incident photon energy, and particularly on angular velocity~\cite{Pereira_2024,quantum6040041}.

The recent developments discussed above build upon the well-known QR model proposed by Tan and Inkson~\cite{W-CTan_1996,PRB.1999.8.5626}. This model enables an analytical treatment of the energy spectrum and wavefunctions, allowing for a detailed analysis of rotation-induced modifications. Studying such systems under external fields and rotational motion provides a fertile ground for exploring fundamental quantum phenomena and potential applications in quantum devices.

In this work, we adopt the Tan–Inkson confinement model and introduce rotational effects via an effective non‑inertial coupling. This framework enables a systematic investigation of how rotation alters three fundamental electronic properties in two-dimensional quantum rings: the Fermi energy, magnetization, and persistent current.

These properties have been extensively examined in low‑dimensional systems. Francisco et al. analyzed the influence of an anisotropic effective mass on electronic structure, magnetization, and persistent currents in a semiconductor quantum ring with conical geometry \cite{PE.2024.158.115898}. Bulaev et al. explored how surface curvature affects the magnetic moment and persistent currents in two‑dimensional rings and quantum dots \cite{PRB.2004.69.1955313}. Additionally, Furtado investigated the interplay of Landau quantization and curvature in two‑dimensional quantum dots \cite{EP.2007.79.57001}. Further significant contributions to the study of Fermi energy, magnetization, and persistent currents in low‑dimensional systems can be found in Refs.\ \cite{FWS.2022.63.58,PRB.2024.110.195426,PRB.2017.96.165407}.

The structure of this article is organized as follows. In Sec. \ref{TImodel}, we present the QR model with rotational effects. In Sec. \ref{FE-M-PC}, we numerically analyze the influence of rotation on the Fermi energy, magnetization, and persistent current. Finally, in Section~\ref{last}, we present our conclusions and final remarks.

\section{The Tan--Inkson Model with Rotation}
\label{TImodel}
In this section, we develop the theoretical model for a two-dimensional quantum ring formed in a GaAs heterostructure, where the transverse confinement is modeled by the Tan-Inkson potential. We extend the conventional treatment by incorporating a uniform rotation into the ring, which manifests itself through Coriolis and centrifugal contributions in the Hamiltonian. This allows us to quantitatively analyze how the rotational dynamics influences the physical observables - the Fermi energy, the magnetization and the persistent current - in the Tan-Inkson quantum ring model.

We consider an electron confined in a two-dimensional ring by the Tan-Inkson potential \cite{SST.1996.11.1635,PRB.1999.8.5626}
\begin{equation}
V(r) = \frac{a_1}{r^2} + a_2 r^2 - V_0,
\end{equation}
where $V_0 = 2\sqrt{a_1 a_2}$ ensures that the minimum of the potential occurs at $r = r_0 = (a_1/a_2)^{1/4}$.

\begin{figure}[tbh]
    \centering
    \includegraphics[width=1\linewidth]{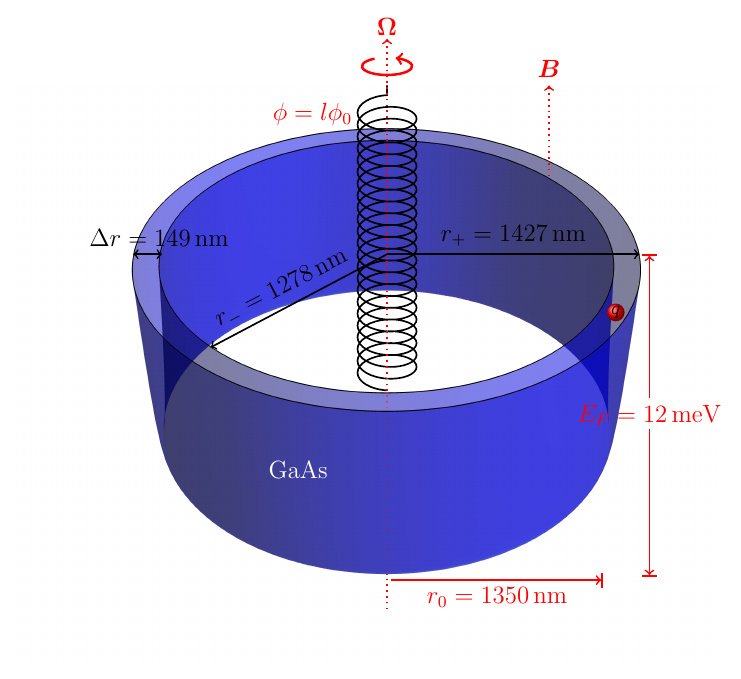}
    \caption{Potential profile of a Tan-Inkson quantum ring with radius \(r_0 = 1350\,\mathrm{nm}\), confinement energy \(\hbar\omega_0 = 2.23\,\mathrm{meV}\), and Fermi energy \(E_F = 12\,\mathrm{meV}\). In the graphic, $q=-e$ represents the charge of the electron.}
    \label{fig:pot}
\end{figure}

Fig.~\ref{fig:pot} shows the Tan-Inson potential for a quantum ring with an average radius of $1350\,\mathrm{nm}$ and parameter $\hbar\omega_{0} = 2.23\,\mathrm{meV}$. The present study is conducted using this configuration, with the Fermi energy fixed at $12\,\mathrm{meV}$. In the figure, a thin solenoid can be seen passing through the center of the ring, representing the AB flux, as well as the coupling with the magnetic field and the vector $\boldsymbol{\Omega}$, indicating rotation around the $z$-axis. The quantum ring under analysis is composed of a GaAs heterostructure.

To include the effects of rotation, we move to a rotating frame with angular velocity $\boldsymbol{\Omega} = \Omega \hat{z}$. The total Hamiltonian of the system, including magnetic and rotational effects, is written as
\begin{equation}
    H = \frac{1}{2\mu} \left( \boldsymbol{p} - e\boldsymbol{A} - \mu \boldsymbol{\Omega} \times \boldsymbol{r} \right)^2 - \frac{1}{2} \mu (\boldsymbol{\Omega} \times \boldsymbol{r})^2 + V(r),
\end{equation}
where $\mu$ is the effective mass and $\boldsymbol{A}$ is the vector potential associated with the external magnetic field and AB flux. In cylindrical coordinates, the effects of these two quantities on the ring are considered separately as
\begin{equation}
    \boldsymbol{A} = \frac{1}{2}Br\hat{\varphi} + \frac{\phi}{2\pi r}\hat{\varphi},
\end{equation}
where $B$ is the magnetic field strength and $\phi=l\phi_0$ is the magnetic flux, and $\phi_0 =h/e$ is the flux quantum.

Using the ansatz $\Psi(r,\varphi) = R(r)e^{im\varphi}$, we obtain the radial Schr\"odinger equation:
\begin{equation}
    R''(r) + \frac{1}{r} R'(r) + \left[ -\frac{L^2}{r^2} - \frac{\mu^2 \omega_{\text{eff}}^2 r^2}{4\hbar^2} + \gamma' \right] R(r) = 0,
\end{equation}
where
\begin{align}
    L^2 &= (m - \phi)^2 + \frac{2\mu a_1}{\hbar^2}, \\
    \omega_{\text{eff}}^2 &= \omega_c^2 + 4\Omega\omega_c + \omega_0^2, \\
    \omega^* &= \omega_c + 2\Omega, \\
    \gamma' &= \frac{\mu \omega^*}{\hbar}(m - \phi) + \frac{2\mu}{\hbar^2}(V_0 + E),
\end{align}
and $\omega_c = eB/\mu$, $\omega_0^2 = 8a_2/\mu$.

The energy eigenvalues are then given by:

\begin{align}
    E_{n,m} = &\left(n + \frac{1}{2} \sqrt{(m - \phi)^2 + \frac{2\mu a_1}{\hbar^2}} + \frac{1}{2} \right) \hbar \omega_{\text{eff}}  \nonumber \\ & - \frac{\hbar}{2}(m - \phi)\omega^* - \frac{\mu}{4} \omega_0^2 r_0^2.\label{eq:energy}
\end{align}

As expected, in the absence of rotation, our results agree with those reported in Ref. \cite{PRB.1999.8.5626}.

\section{Fermi Energy, Magnetization, and Persistent Current}
\label{FE-M-PC}

We now proceed to analyze how rotation influences the system's physical properties: Fermi energy, magnetization, and persistent current. All numerical results presented below are calculated under the assumption of zero temperature and $\phi = 0$.

Starting from the analytical form of the energy spectrum in Eq. \eqref{eq:energy}, we can investigate the Fermi energy, which is determined by the total number of electrons and the energy levels below the Fermi surface.  

The statistical physics reveals that the Fermi-Dirac distribution function is given by \cite{landau2013statistical,kittel1980thermal}
\begin{equation}
f\left(E_{n, m}\right)=\frac{1}{1+e^{\left(E_{n, m}-\mu_c\right) / k_B T}},
\end{equation}
where $\mu_c = \mu_c(T)$ is the temperature-dependent chemical potential. In the limit as $T \to 0$, we have $\mu_c=E_F$. with $E_{F}$ representing the Fermi energy. In this case, the distribution takes on a step-like form: $f\left(E_{n, m}\right)=1$ for $E_{n, m}<E_{F}$ and $f\left(E_{n, m}\right)=0$ for $E_{n, m}>E_{F}$. Thus, the total number of electrons $N_{e}$ is obtained by integrating $f\left(E_{n, m}\right)$ over all available states, so that \cite{AdP.2019.531.1900254}
\begin{equation}
    N_{e}=\sum_{n,m}\theta\left(E_{F}-E_{n,m}\right),
\end{equation}
where  $\theta\left(E_{F}-E_{n,m}\right)$ is the Heaviside function.

\begin{figure}[htbp]
\centering
\includegraphics[width=1.0\linewidth]{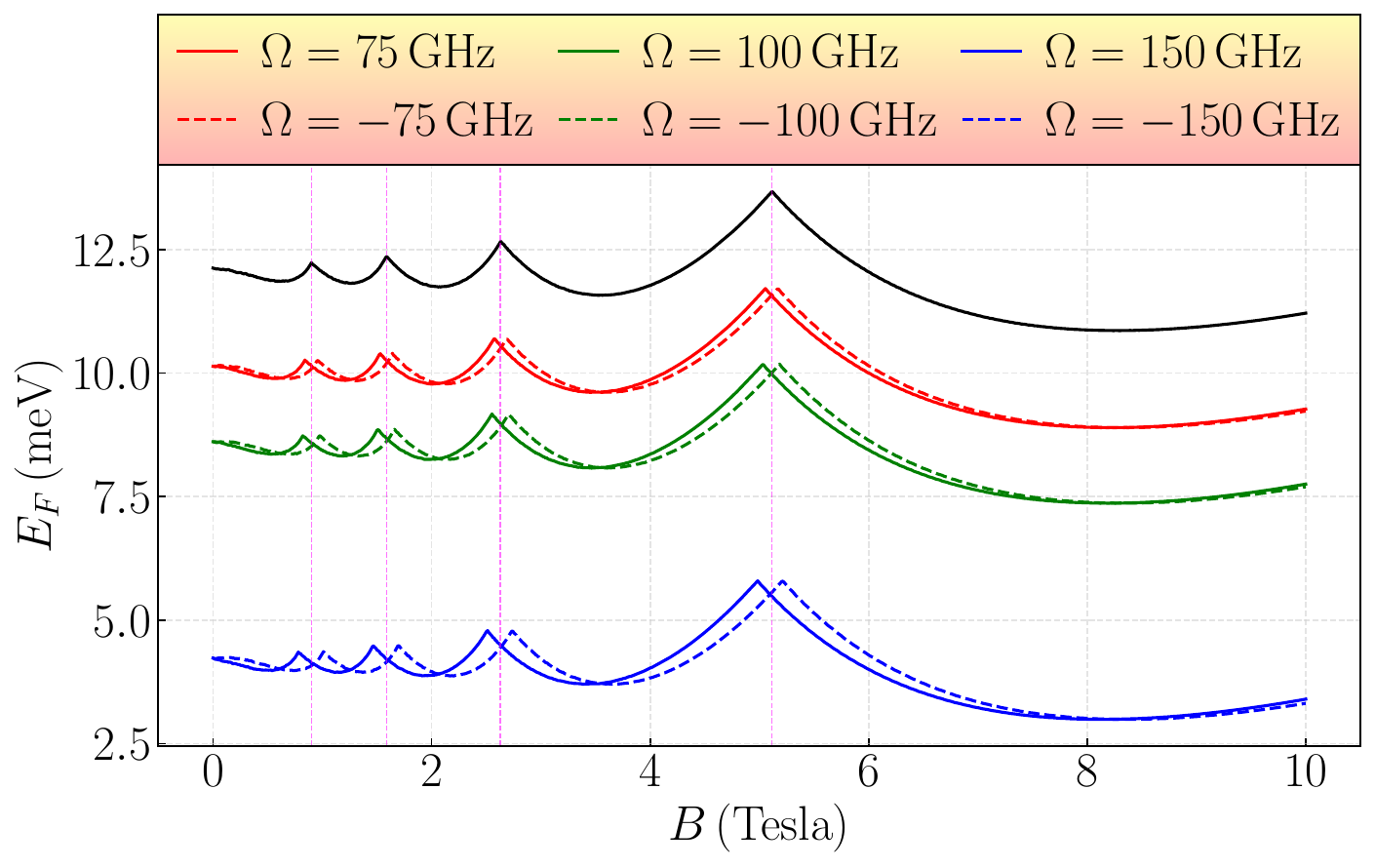}
\caption{The zero-temperature Fermi energy as a function of magnetic field for different values of $\Omega$. The continuous and dashed curves represent the positive and negative rotations, respectively. The black line corresponds to the case without rotation. The parameter values used were $r_0 = 1350\,\mathrm{nm}$, $\hbar\omega_0 = 2.23\,\mathrm{meV}$ and $N_e = 1400$.}
\label{fig:Efermi}
\end{figure}

For a fixed number of electrons, the Fermi energy as a function of the magnetic field is influenced by the rotation present in the quantum ring. In Fig.~\ref{fig:Efermi}, we present the Fermi energy for different values of the rotation magnitude $|\Omega|$ on the order of GHz. It is observed that, as $|\Omega|$ increases (decreases), the Fermi energy curves undergo a reduction (elevation). Additionally, it is noted that the respective peaks of these curves are shifted to lower (higher) magnetic field values as $|\Omega|$ increases (decreases). We also highlight the positive and negative values of $\Omega$: continuous curves represent positive rotation values, while dashed curves correspond to negative rotation values. The oscillations observed in the Fermi energy curves are due to the depopulation of an $n$ sub-band \cite{PE.2021.132.114760}. As the magnitude of $|\Omega|$ increases, the energy minima of the states decrease and, consequently, there is a reduction in the Fermi energy. With the total number of electrons constant, depopulation occurs: the Fermi energy, $E_{F}\approx12\, \text{meV}$ at $B=0$, moves from the $n=4$ subband to $n=3$, then to $n=2$ and finally, to $n=1$, in that order.

Curves of the same color indicate the same absolute value of rotation. It can be seen that as this modulus increases, the separation between the peaks of the same-colored curves also increases. These peaks remain separate for $B\neq0$ and coincide for $B=0$, a direct consequence of the cyclotron frequency. In addition, we see that, for a particular curve, the Fermi energy peaks increase with $B$, reaching their highest value in the range between $B = 4$ and $6~\text{(Tesla)}$.
The vertical magenta lines are centered on the values of $B$ at which the Fermi energy peaks reach their maximum value, serving as a reference for quantifying the shifts of the other curves as $|\Omega|$ increases.

\begin{figure}[tbhp]
\centering
\includegraphics[width=1.0\linewidth]{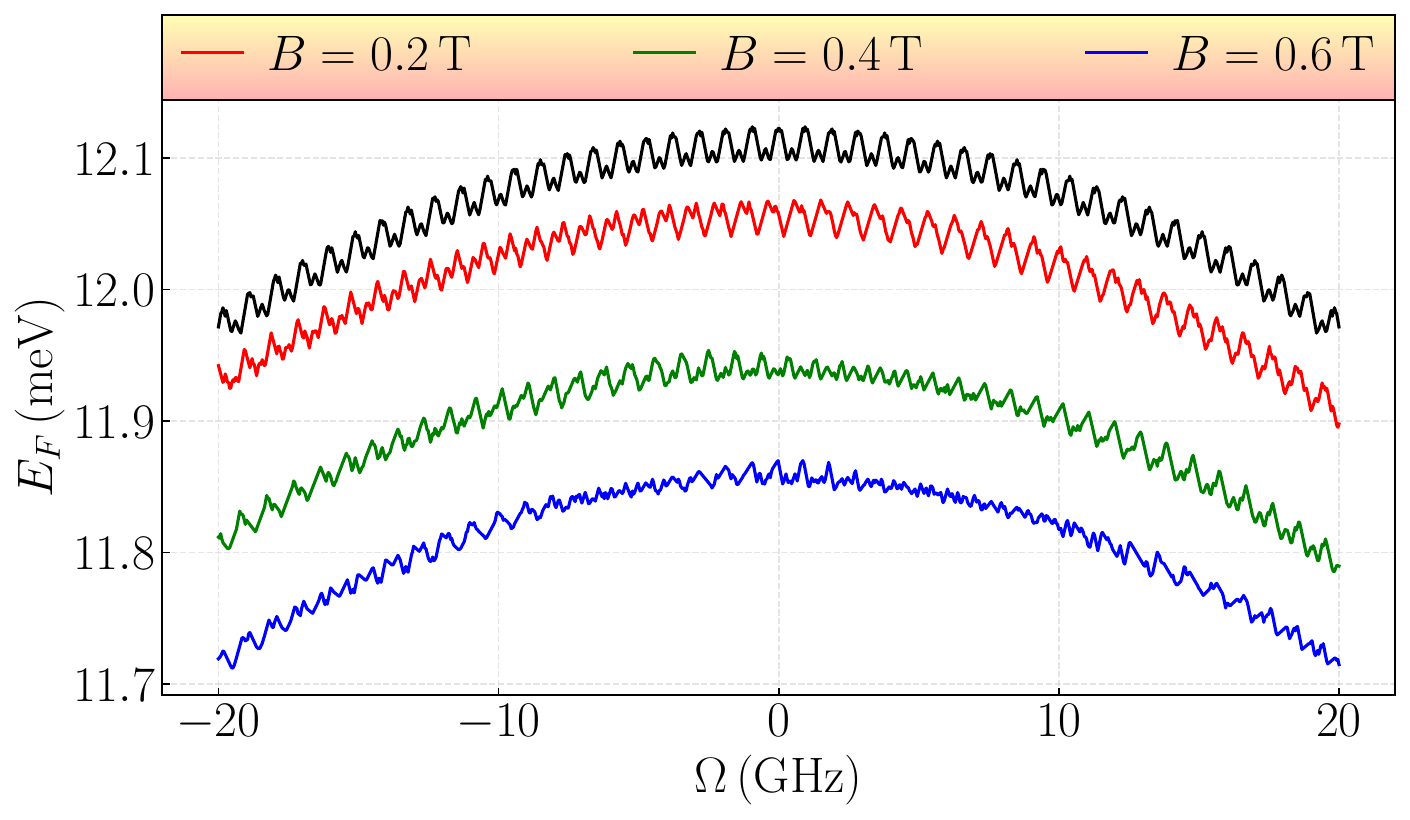}
\caption{Variation of the Fermi energy as a function of the rotation \(\Omega\) for different magnetic field strengths. The black line corresponds to the case of zero magnetic field. The parameter values used were $r_0 = 1350\,\mathrm{nm}$, $\hbar\omega_0 = 2.23\,\mathrm{meV}$ and $N_e = 1400$.}
\label{fig:Efermi-omega}
\end{figure}

Fig.~\ref{fig:Efermi-omega} shows the Fermi energy as a function of rotation (in the GHz range) for different magnetic field strengths. It can be seen that, in the absence of an external magnetic field ($B = 0\,\mathrm{T}$), both the amplitude and the oscillations of $E_F$ follow a well-defined pattern as $|\Omega|$ increases. For non-zero values of $B$, the regularity of these oscillations is lost, reflecting the interaction between the cyclotron frequency $\omega_c$ and the rotational term, represented by the effective frequency $\omega_{\mathrm{eff}}$. This irregular behavior of $E_F$ for $B \neq 0$ results from both the change in the degeneracy of the Landau levels and the rotational coupling via $\omega_{\mathrm{eff}}$. Additionally, there is a decrease in the average Fermi energy as the magnetic field increases.
\begin{figure}[tbhp]
\centering
\includegraphics[width=1.0\linewidth]{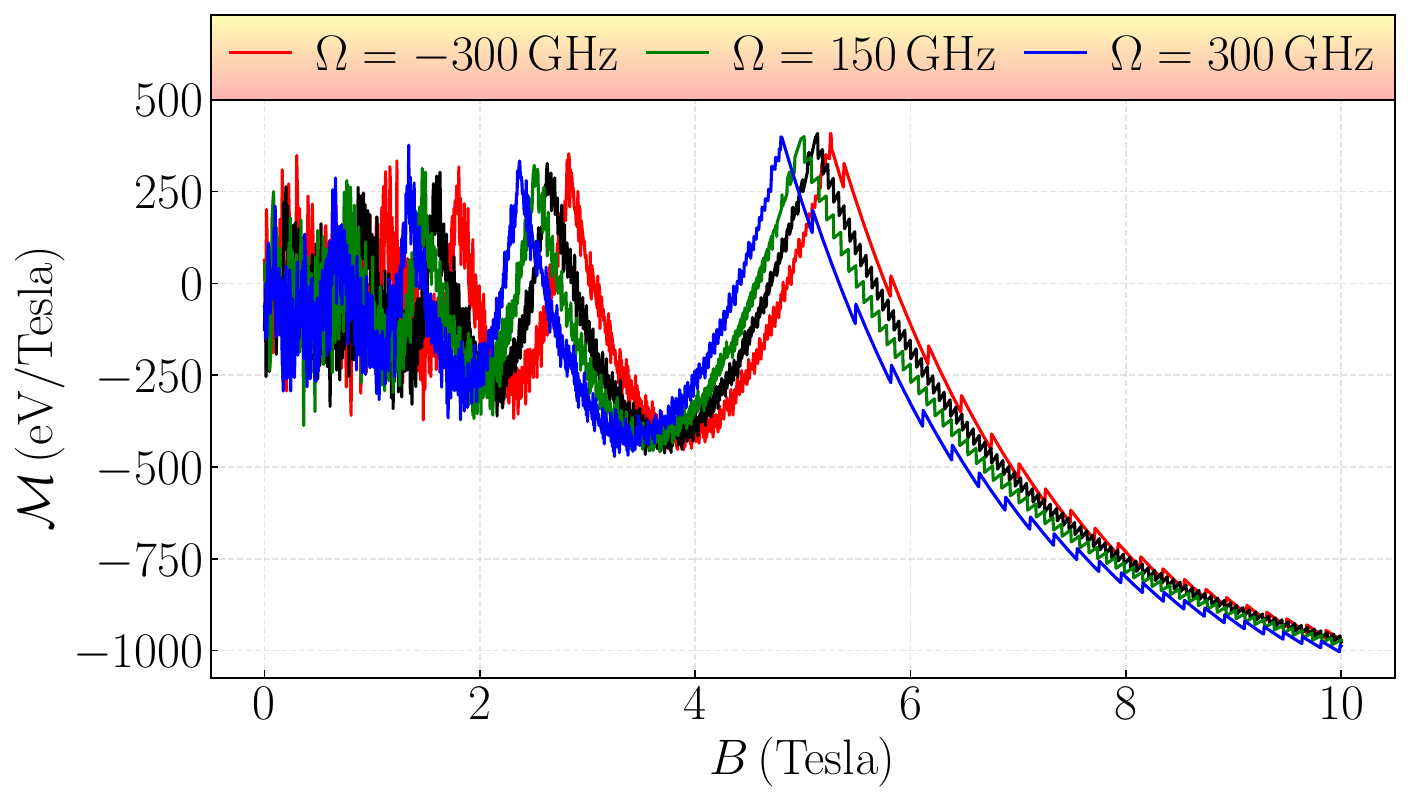}
\caption{Plot of the change in magnetization caused by rotation as a function of the \(B\) magnetic field for different values of $\Omega$. The black line corresponds to the case without rotation. The parameter values used were $r_0 = 1350\,\mathrm{nm}$, $\hbar\omega_0 = 2.23\,\mathrm{meV}$ and $N_e = 1400$.}
\label{fig:magn}
\end{figure}
\begin{figure*}[tbhp]
\centering
\includegraphics[width=0.48\linewidth]{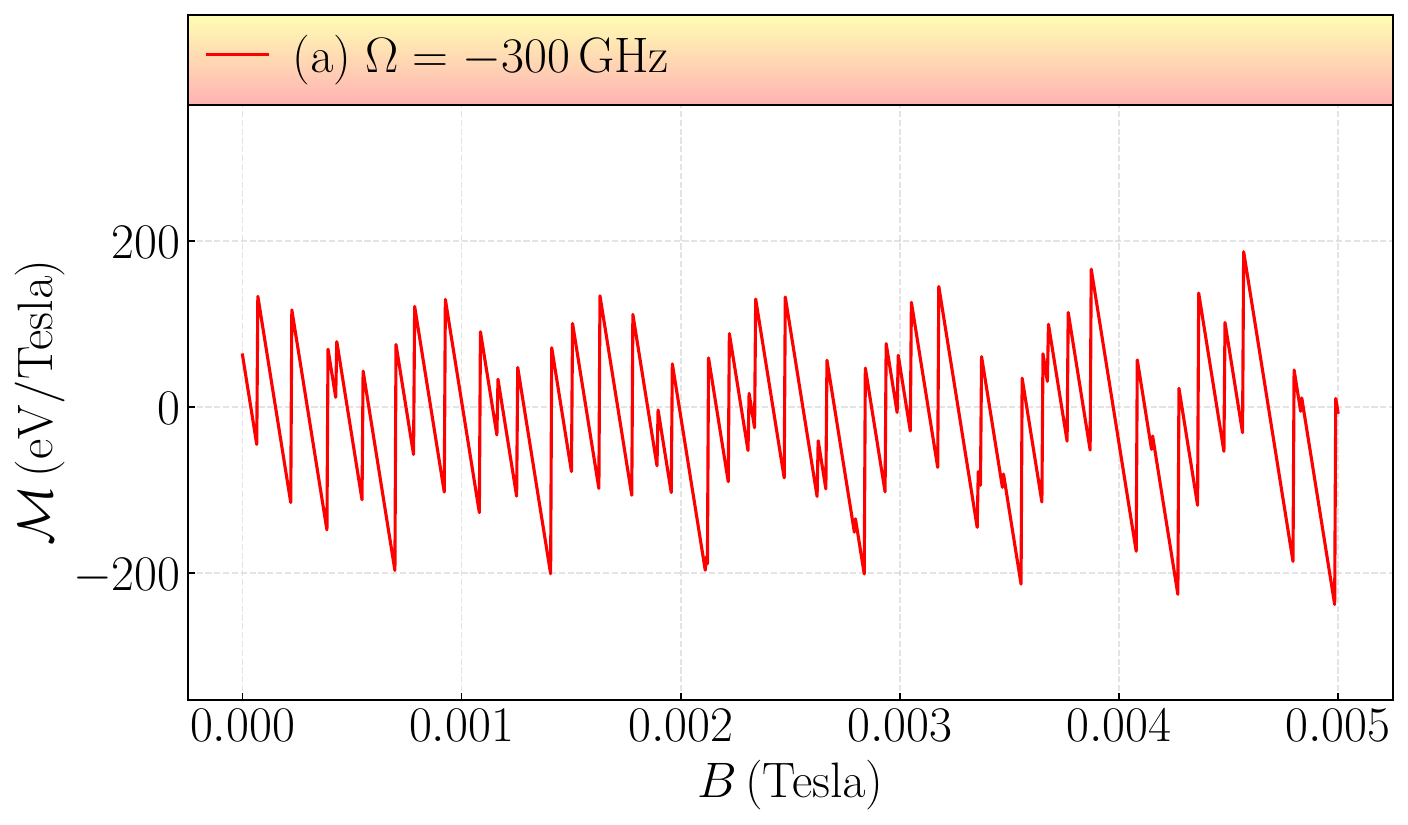}
\includegraphics[width=0.48\linewidth]{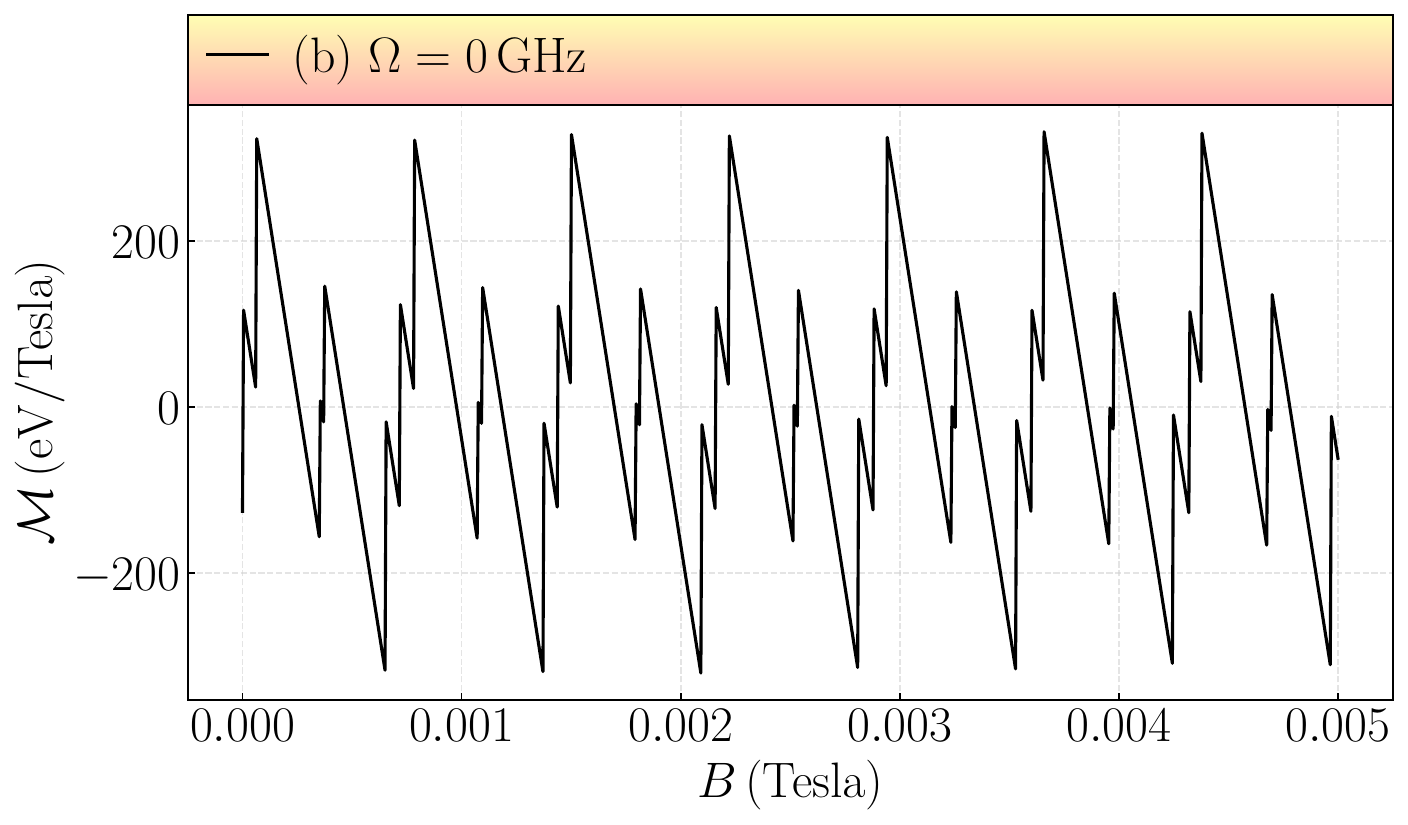}
\includegraphics[width=0.48\linewidth]{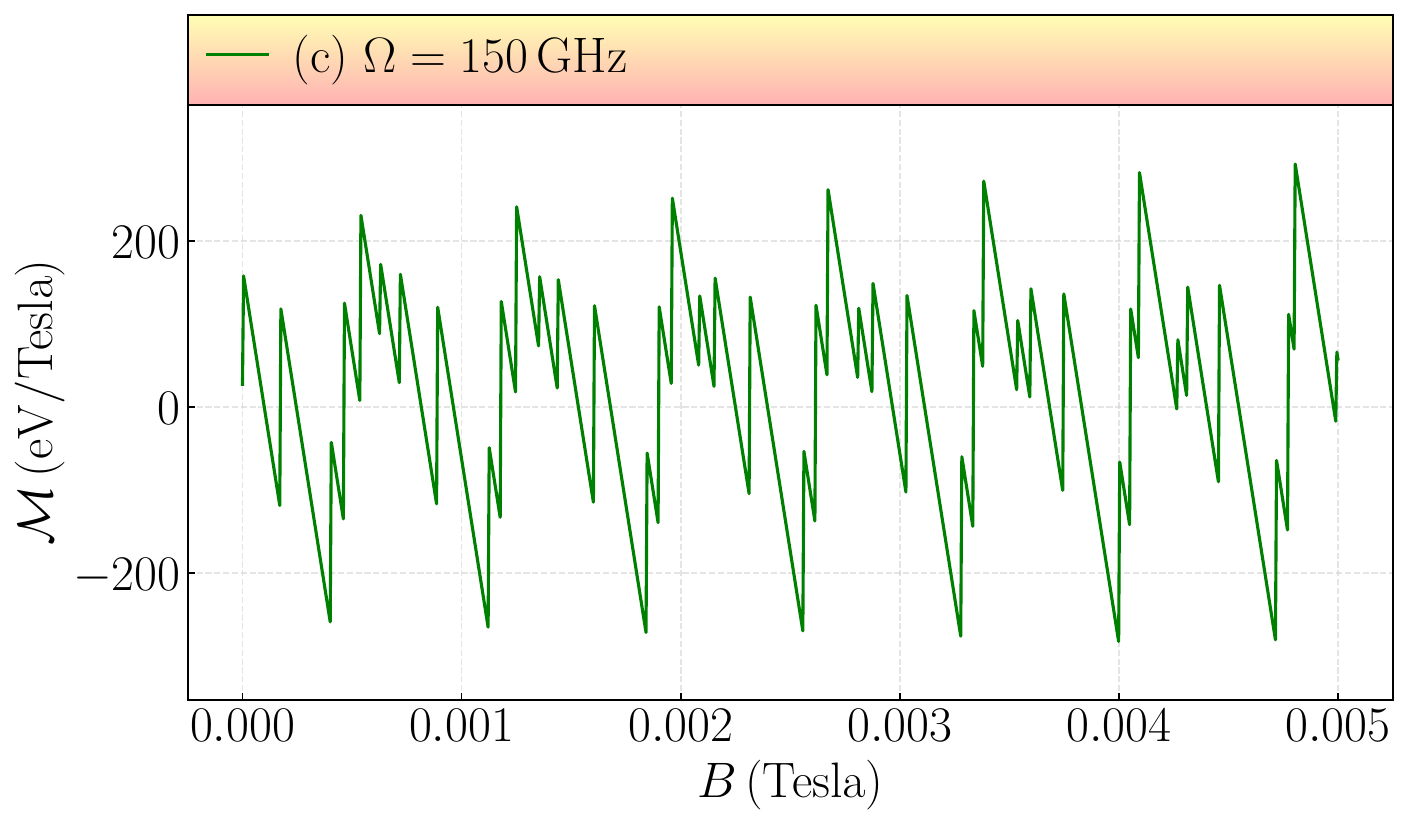}
\includegraphics[width=0.48\linewidth]{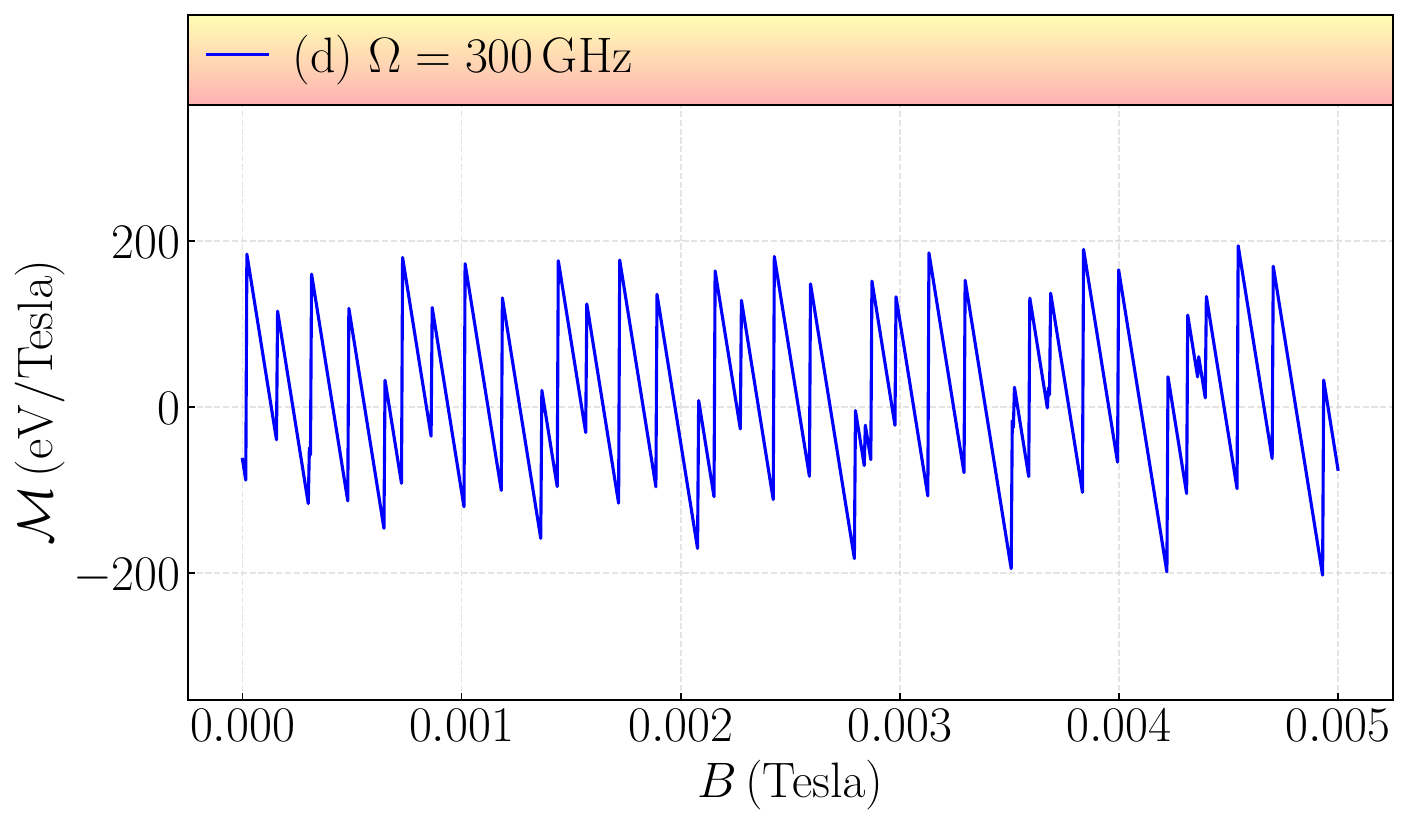}
\caption{Magnetization in the weak magnetic field regime for different values of $\Omega$. The parameter values used were $r_0 = 1350\,\mathrm{nm}$, $\hbar\omega_0 = 2.23\,\mathrm{meV}$ and $N_e = 1400$.}
\label{fig:magnetiw}
\end{figure*}
\begin{figure}[tbhp]
\centering
\includegraphics[width=1.0\linewidth]{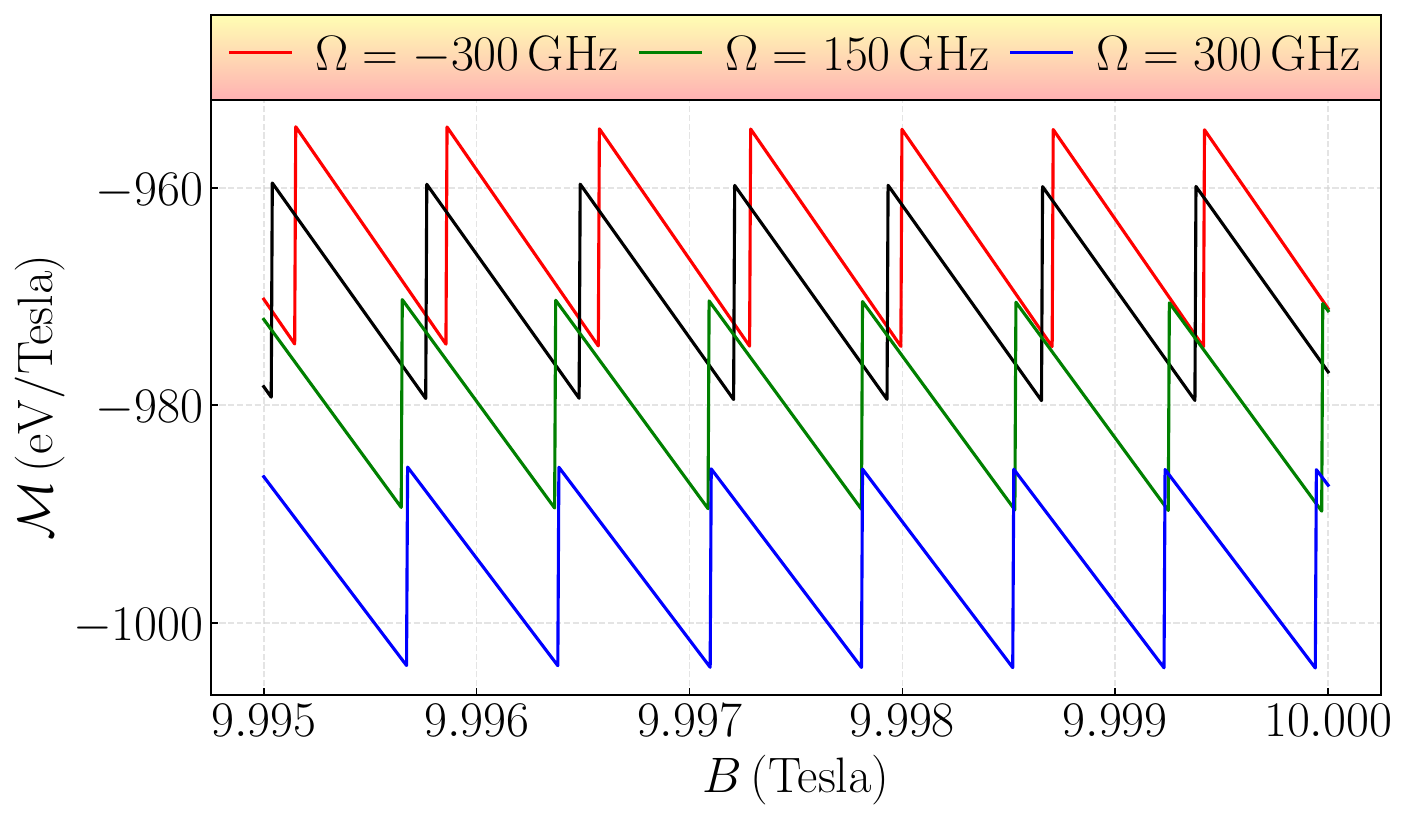}
\caption{Magnetization in the strong magnetic field regime for different values of $\Omega$. The parameter values used were $r_0 = 1350\,\mathrm{nm}$, $\hbar\omega_0 = 2.23\,\mathrm{meV}$ and $N_e = 1400$.}
\label{fig:magns}
\end{figure}
\begin{figure}[tbhp]
\centering
\includegraphics[width=1.0\linewidth]{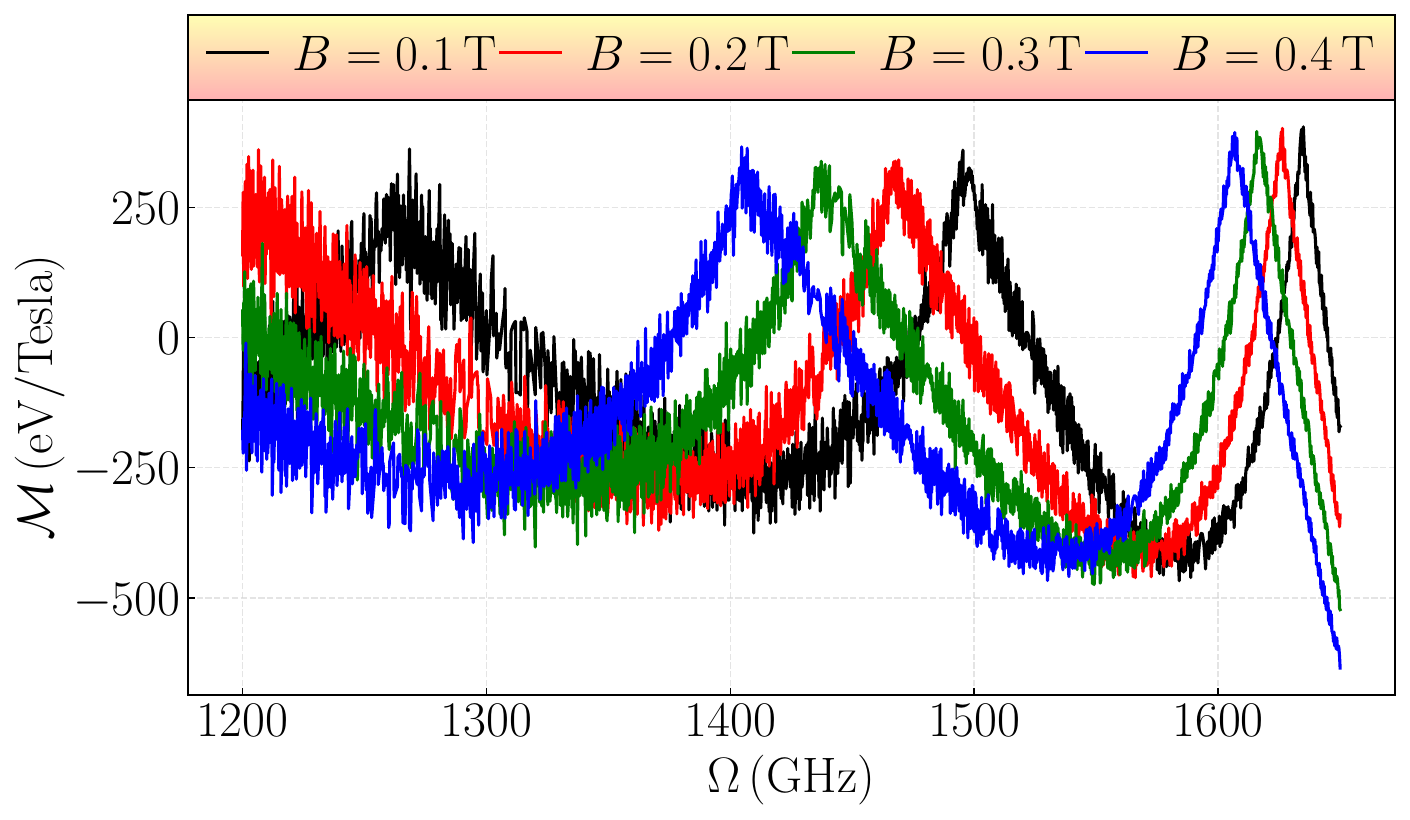}
\caption{Plot of the change in magnetization caused by magnetic field as a function of the \(\Omega\) rotation for different values of $B$. The parameter values used were $r_0 = 1350\,\mathrm{nm}$, $\hbar\omega_0 = 2.23\,\mathrm{meV}$ and $N_e = 1400$.}
\label{fig:magnO}
\end{figure}
\begin{figure*}[tbhp]
\begin{center}
\includegraphics[width=0.48\linewidth]{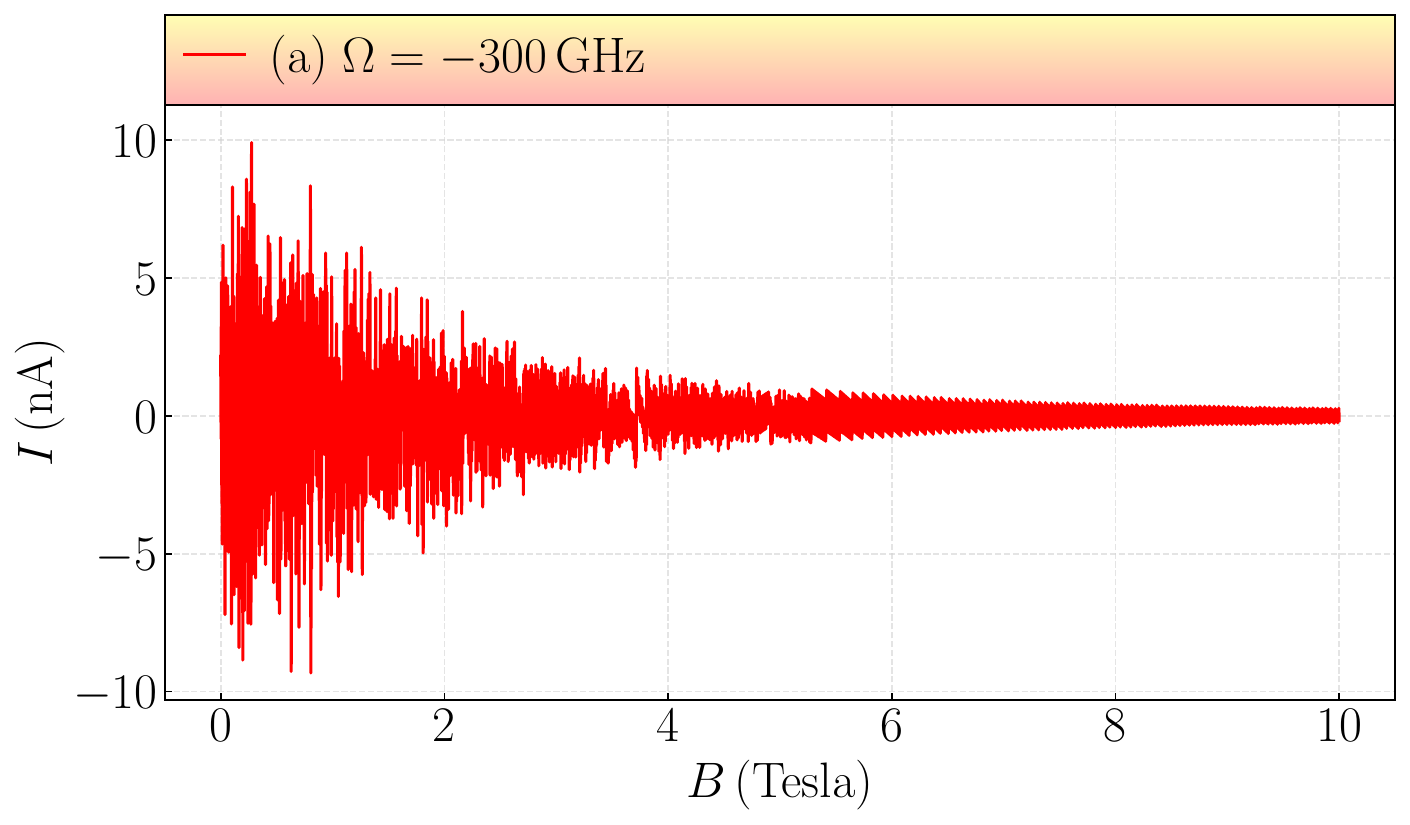}
\includegraphics[width=0.48\linewidth]{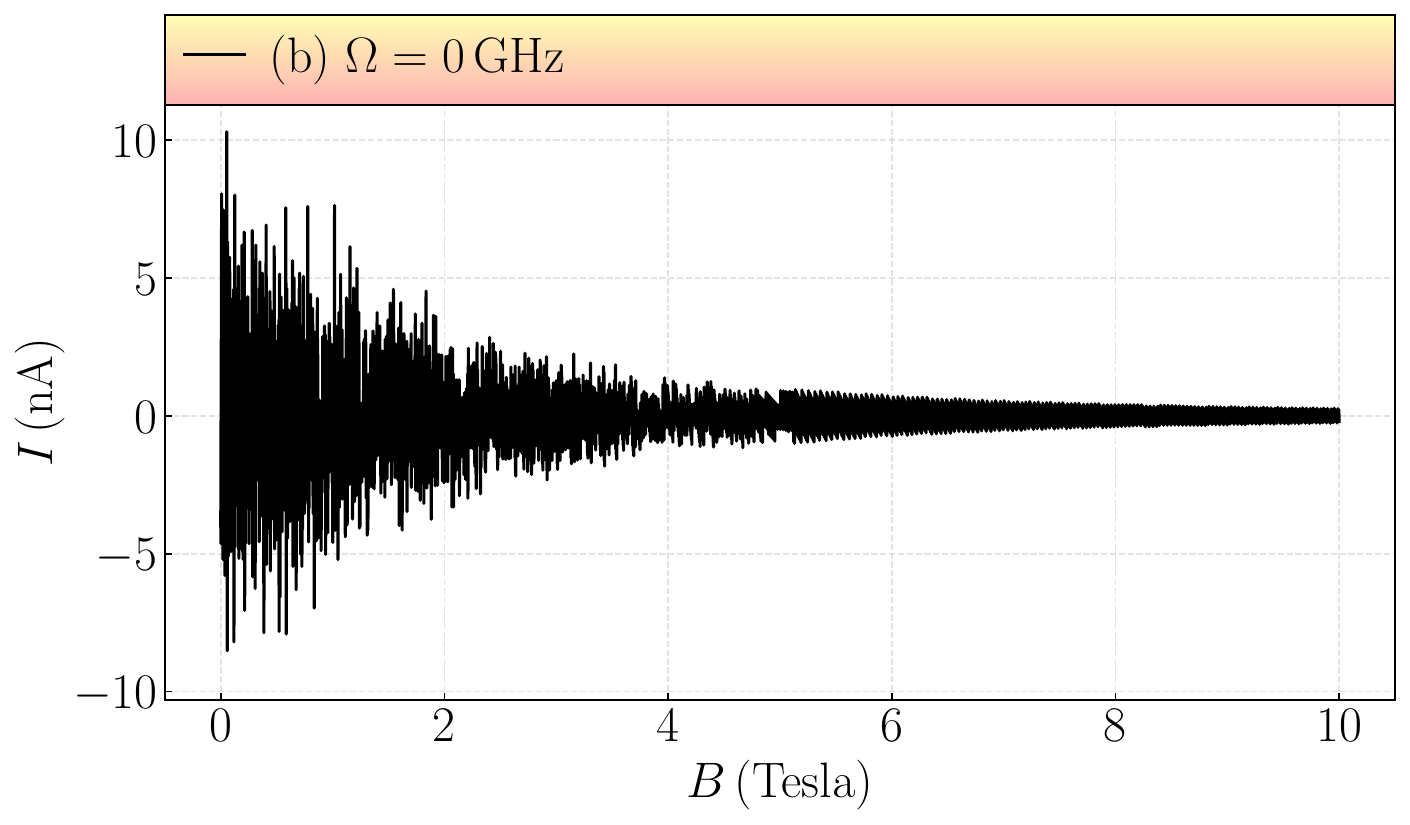}
\includegraphics[width=0.48\linewidth]{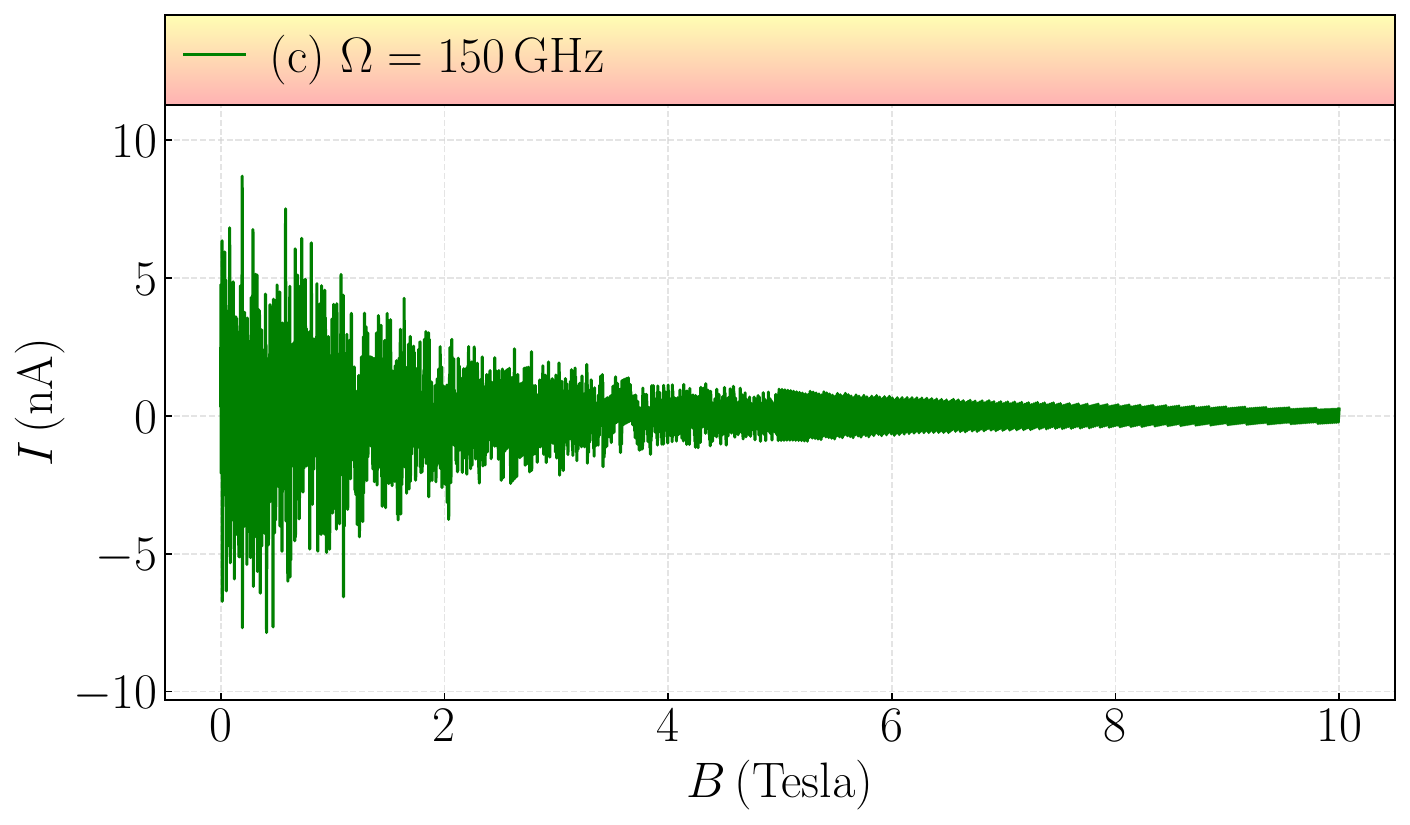}
\includegraphics[width=0.48\linewidth]{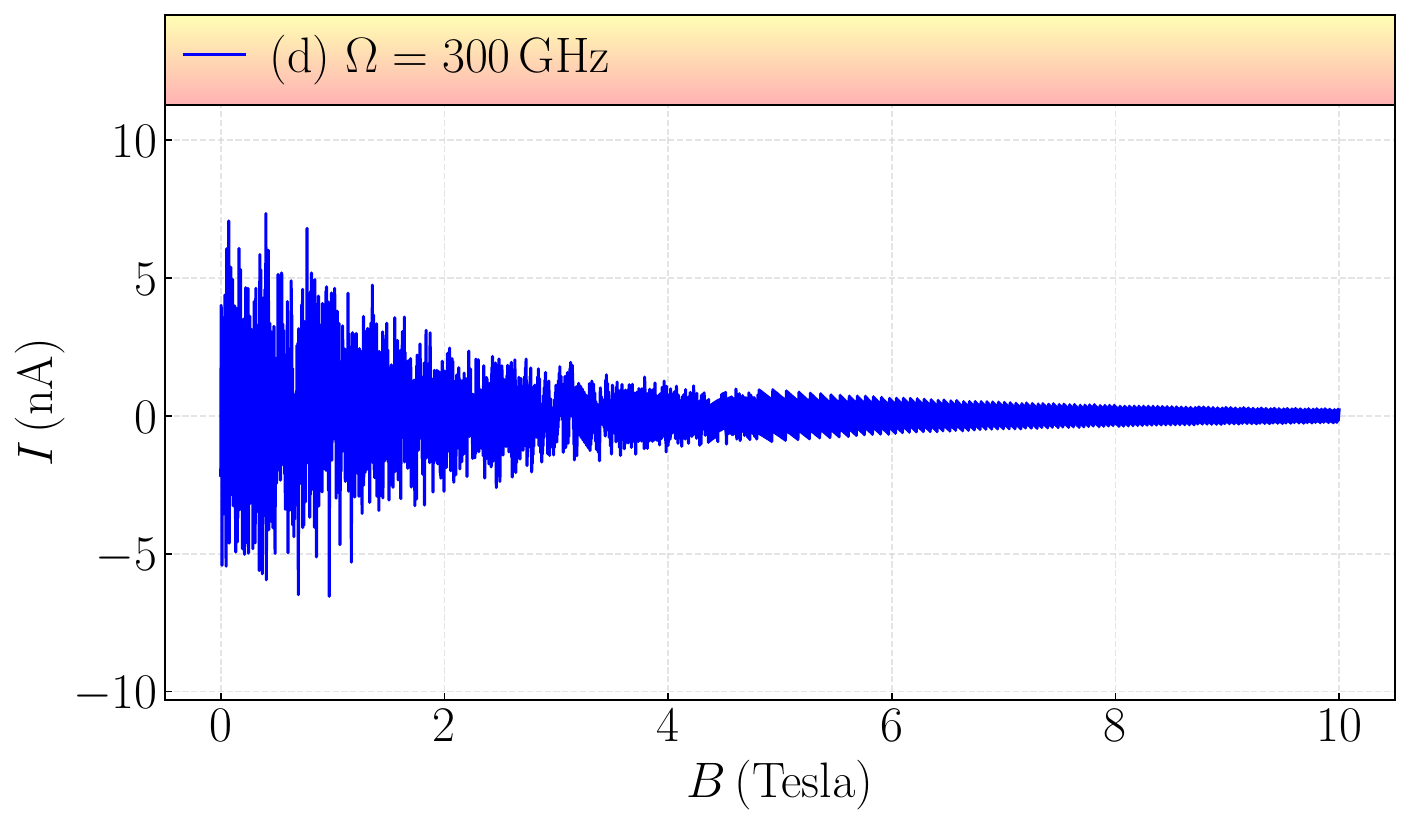}
\caption{Persistent current as a function of the magnetic field for different values of $\Omega$. The parameter values used were $r_0 = 1350\,\mathrm{nm}$, $\hbar\omega_0 = 2.23\,\mathrm{meV}$ and $N_e = 1400$.}
\label{fig:curr}
\end{center}
\end{figure*}
\begin{figure*}[tbhp]
\begin{center}
\includegraphics[width=0.48\linewidth]{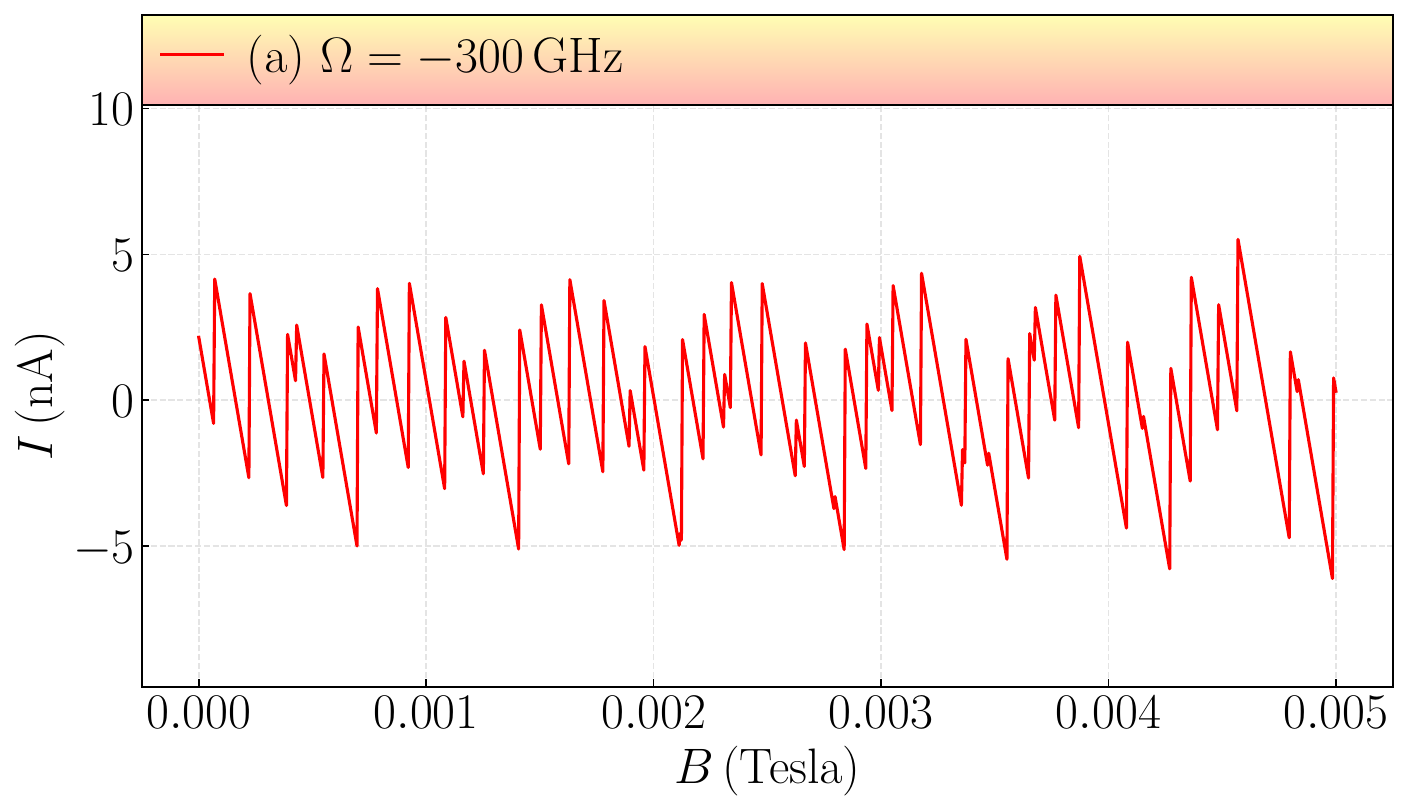}
\includegraphics[width=0.48\linewidth]{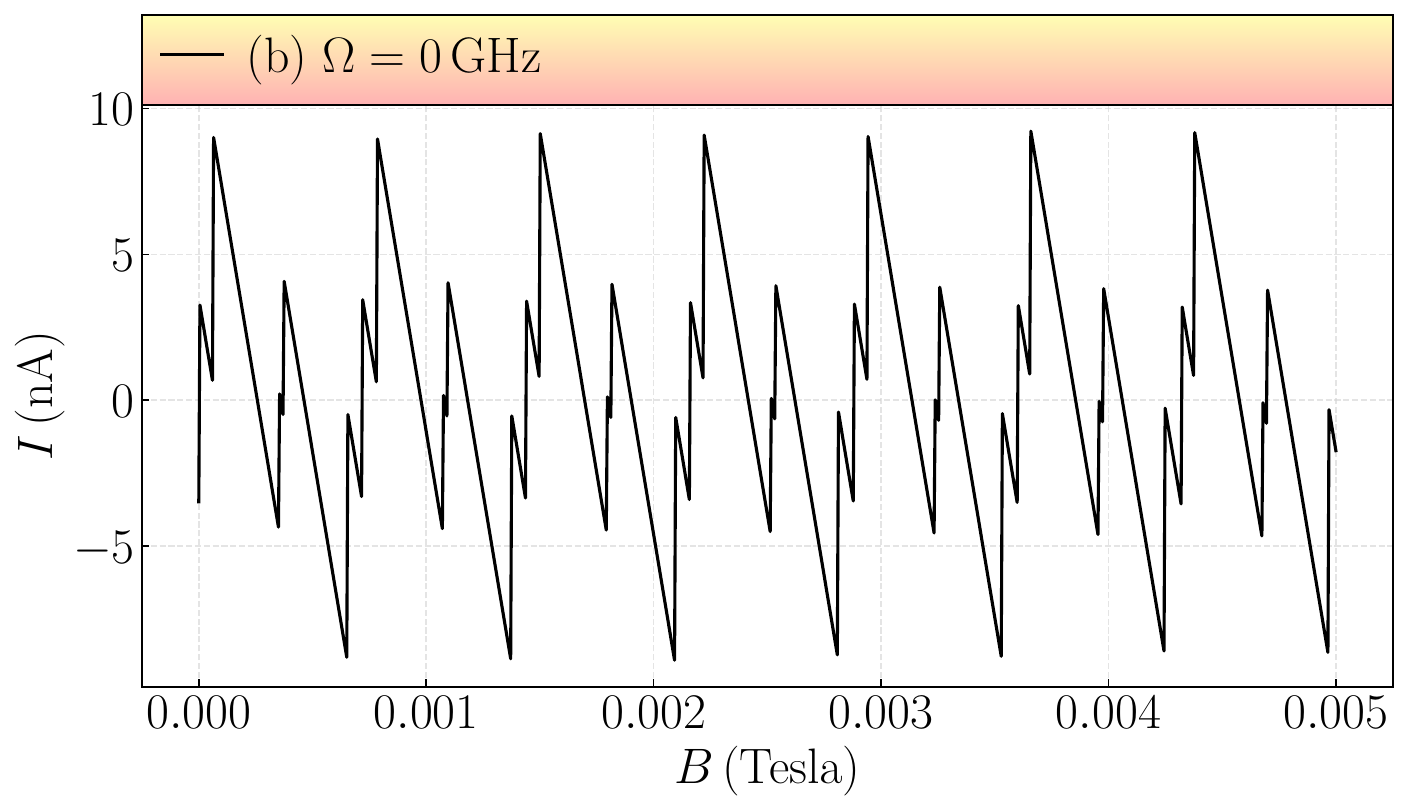}
\includegraphics[width=0.48\linewidth]{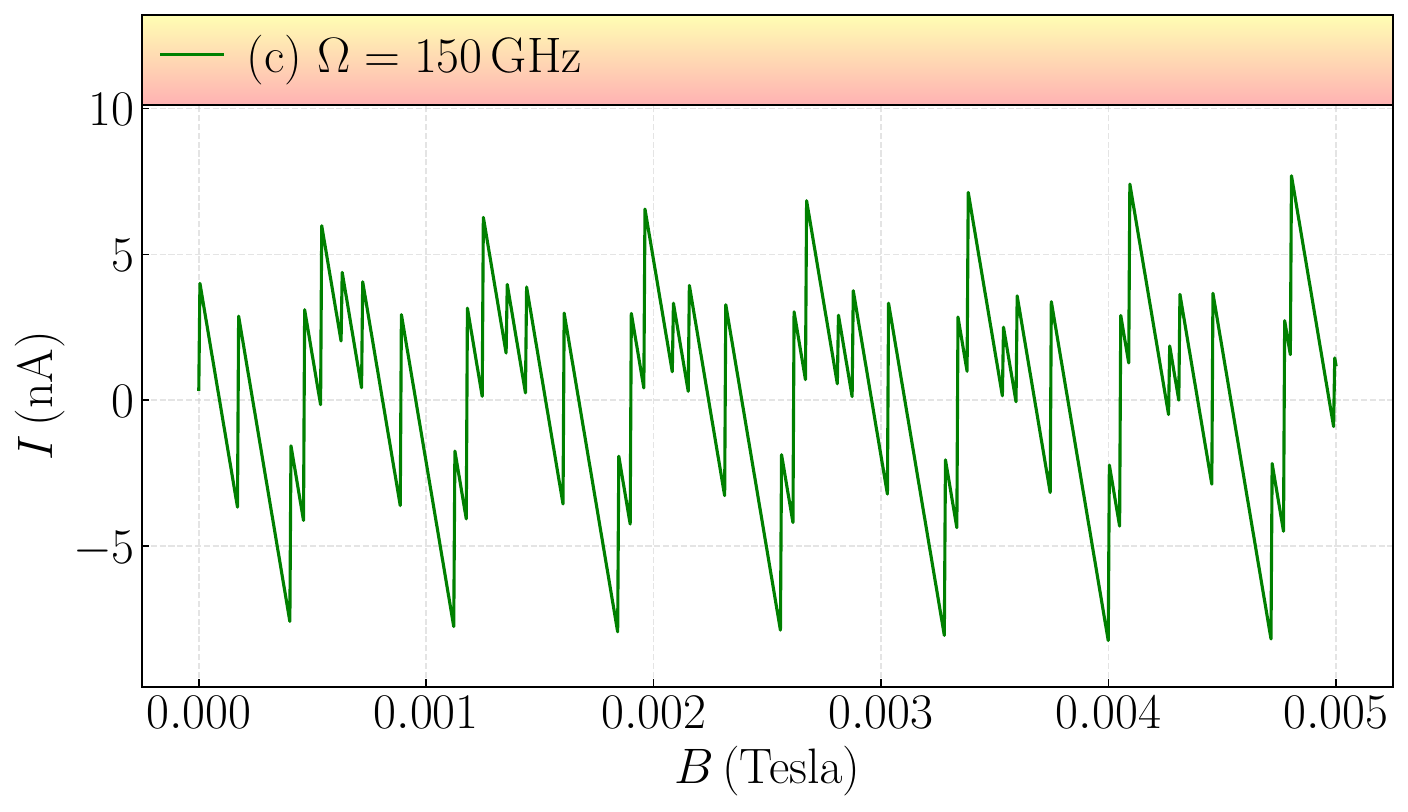}
\includegraphics[width=0.48\linewidth]{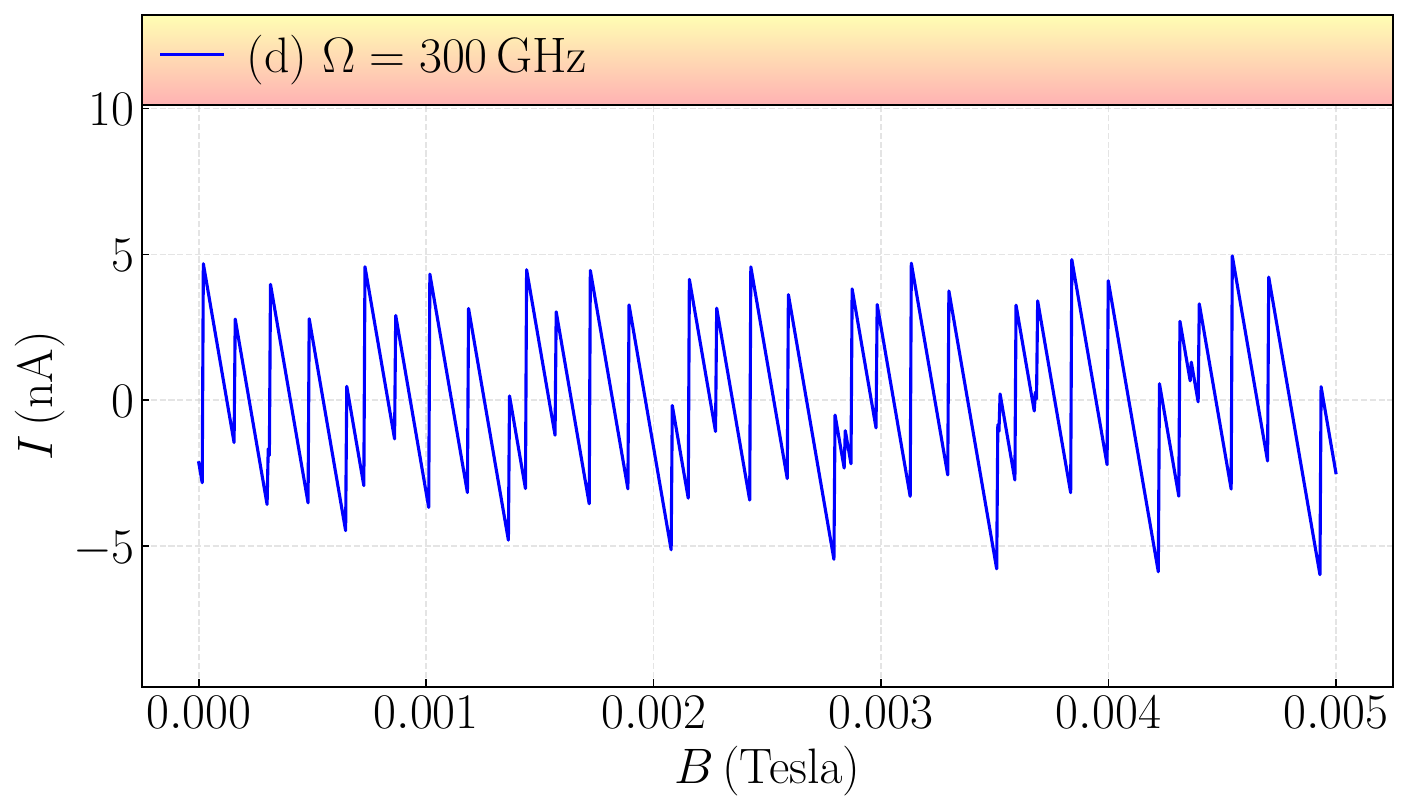}
\caption{Persistent current in the regime of weak magnetic fields for different values of $\Omega$. The parameter values used were $r_0 = 1350\,\mathrm{nm}$, $\hbar\omega_0 = 2.23\,\mathrm{meV}$ and $N_e = 1400$.}
\label{fig:currw}
\end{center}
\end{figure*}
\begin{figure}[tbhp]
\centering
\includegraphics[width=1.0\linewidth]{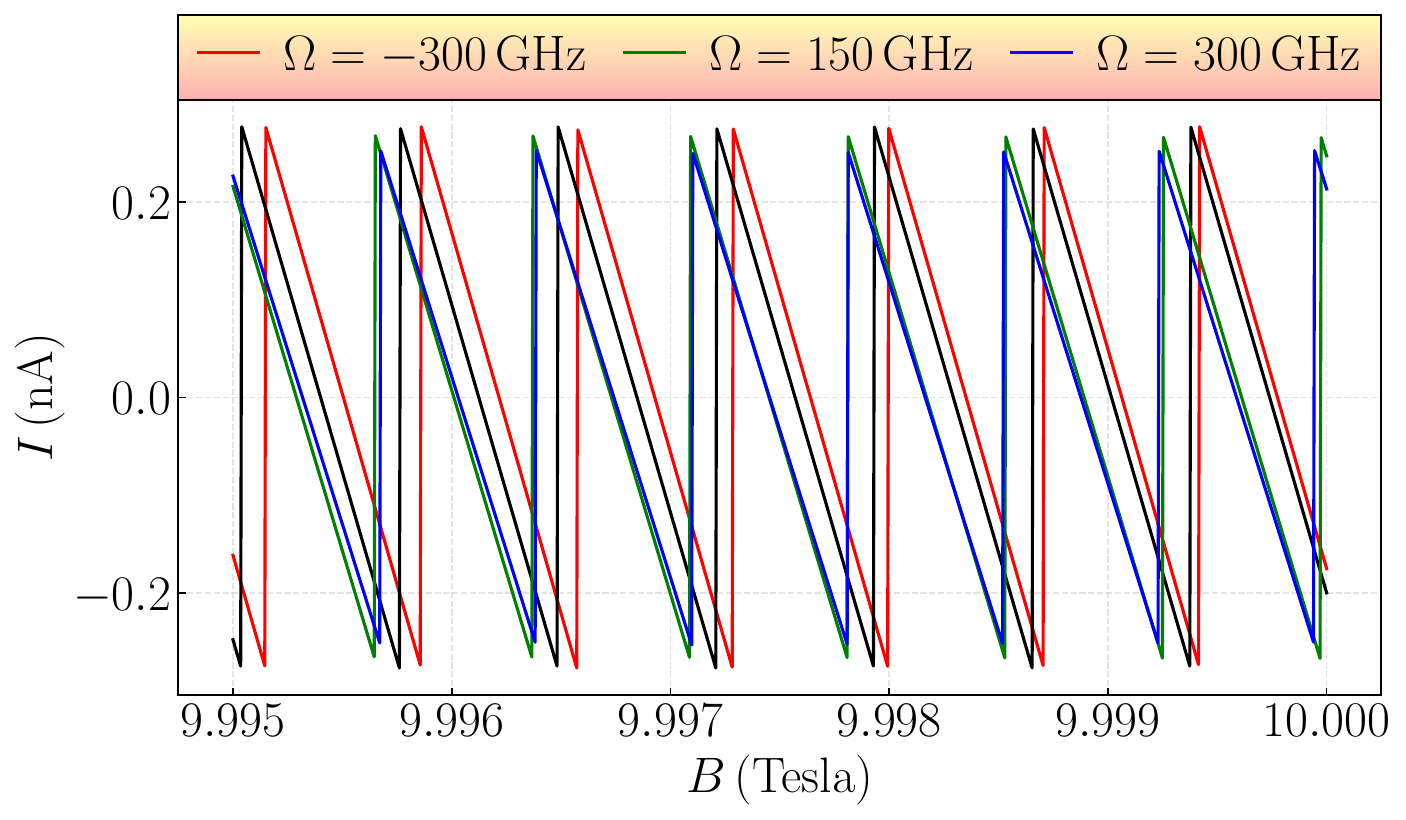}
\caption{Persistent current in the regime of strong magnetic fields for different values of $\Omega$. The parameter values used were $r_0 = 1350\,\mathrm{nm}$, $\hbar\omega_0 = 2.23\,\mathrm{meV}$ and $N_e = 1400$.}
\label{fig:currs}
\end{figure}
\begin{figure*}[tbhp]
\begin{center}
\includegraphics[width=0.48\linewidth]{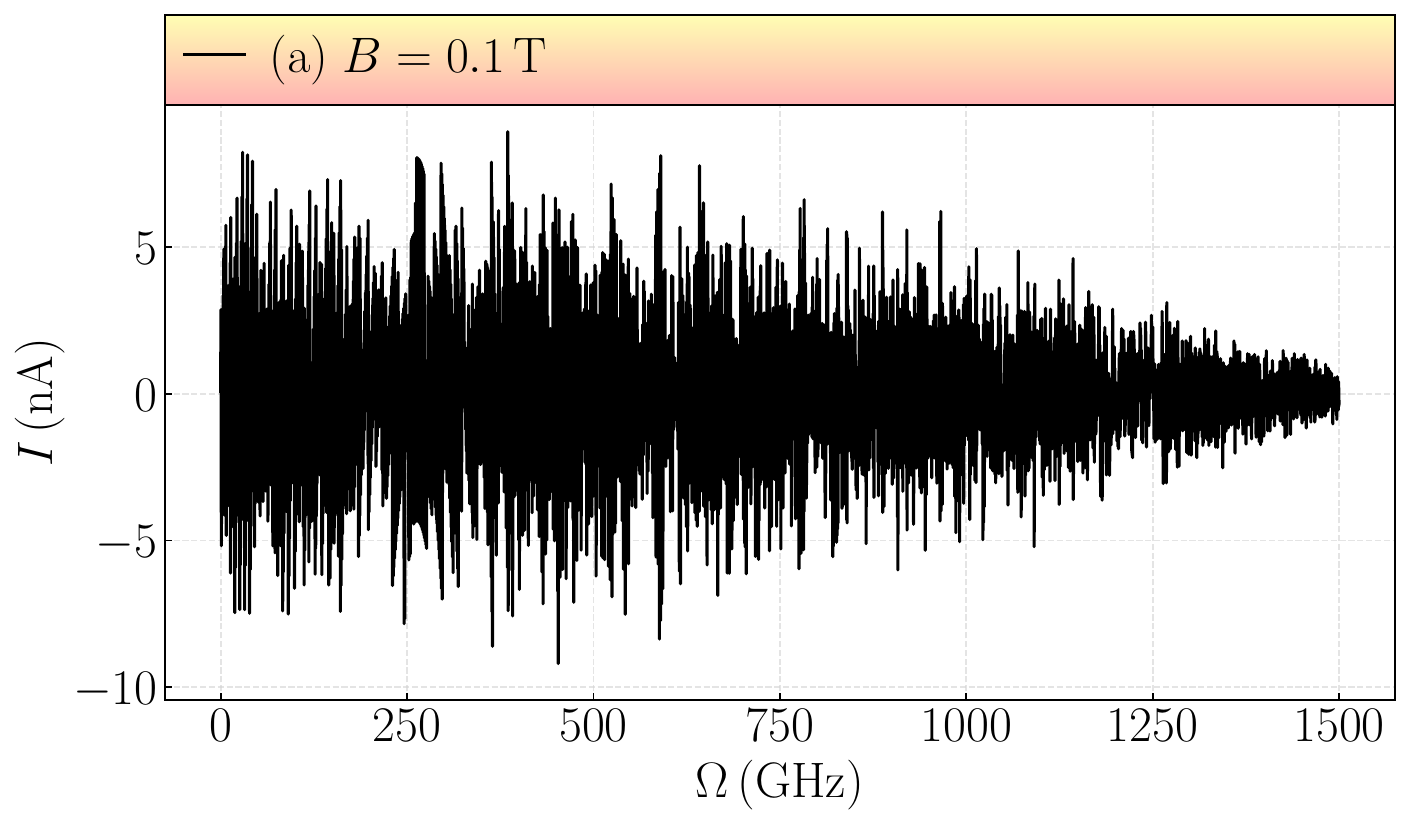}
\includegraphics[width=0.48\linewidth]{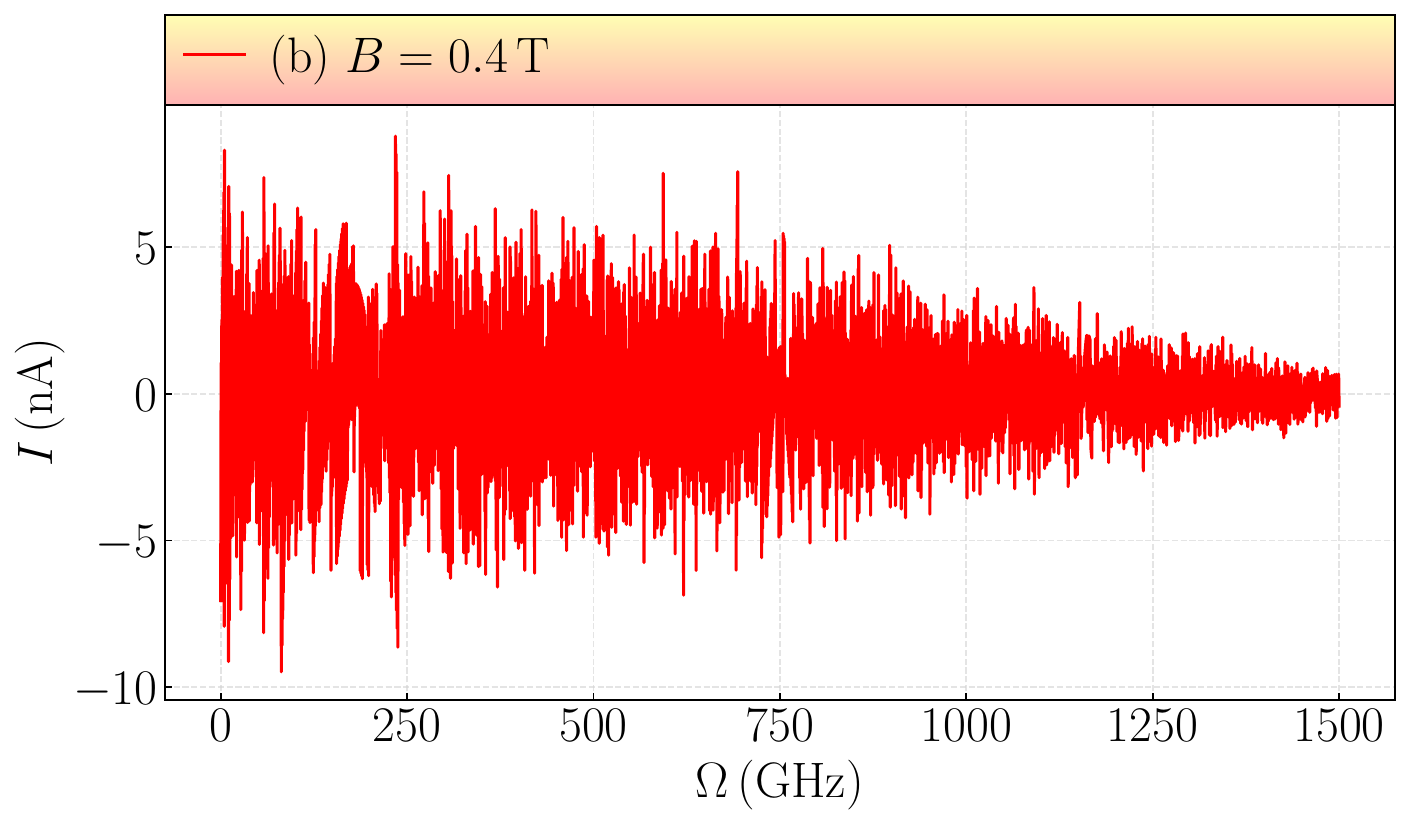}
\includegraphics[width=0.48\linewidth]{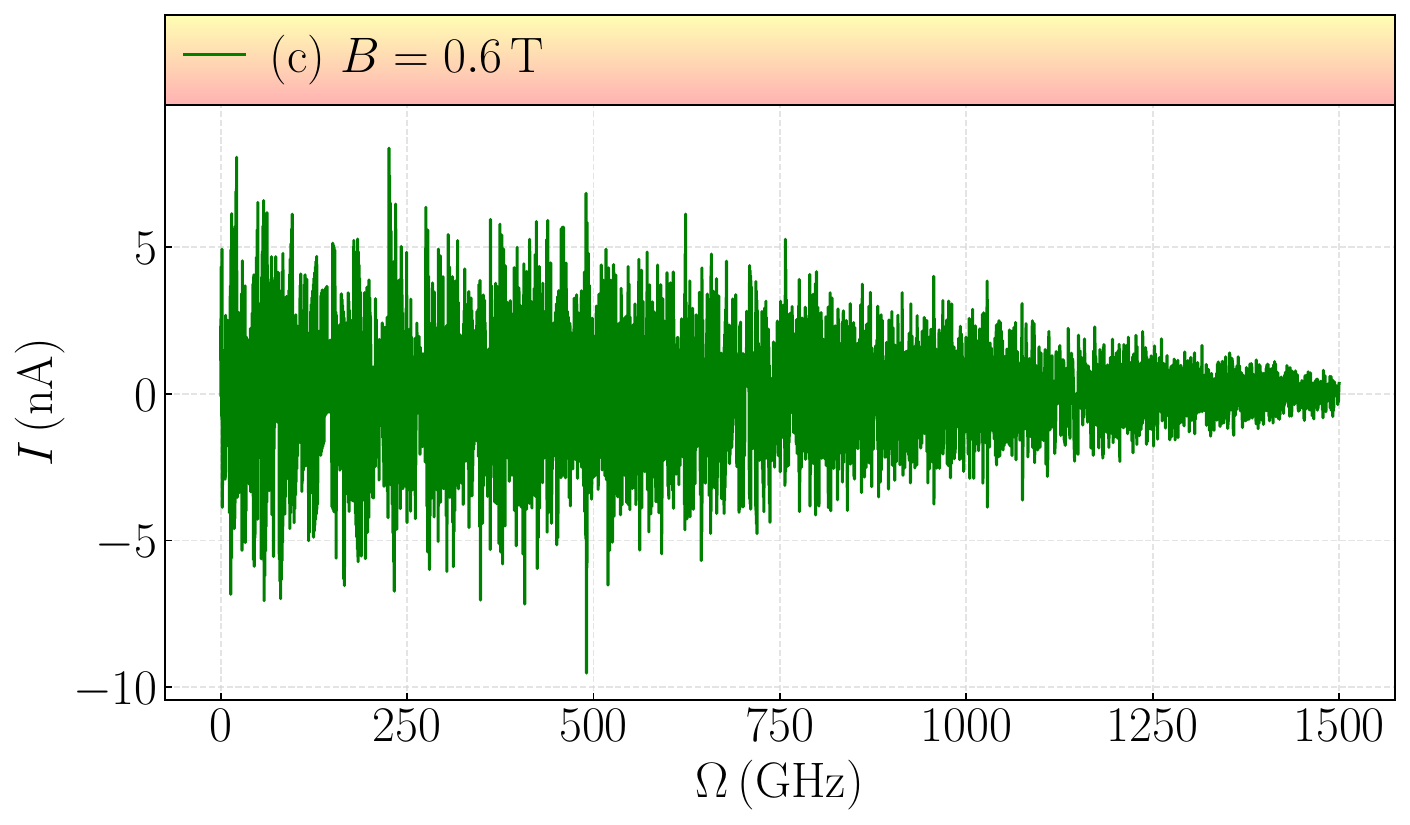}
\includegraphics[width=0.48\linewidth]{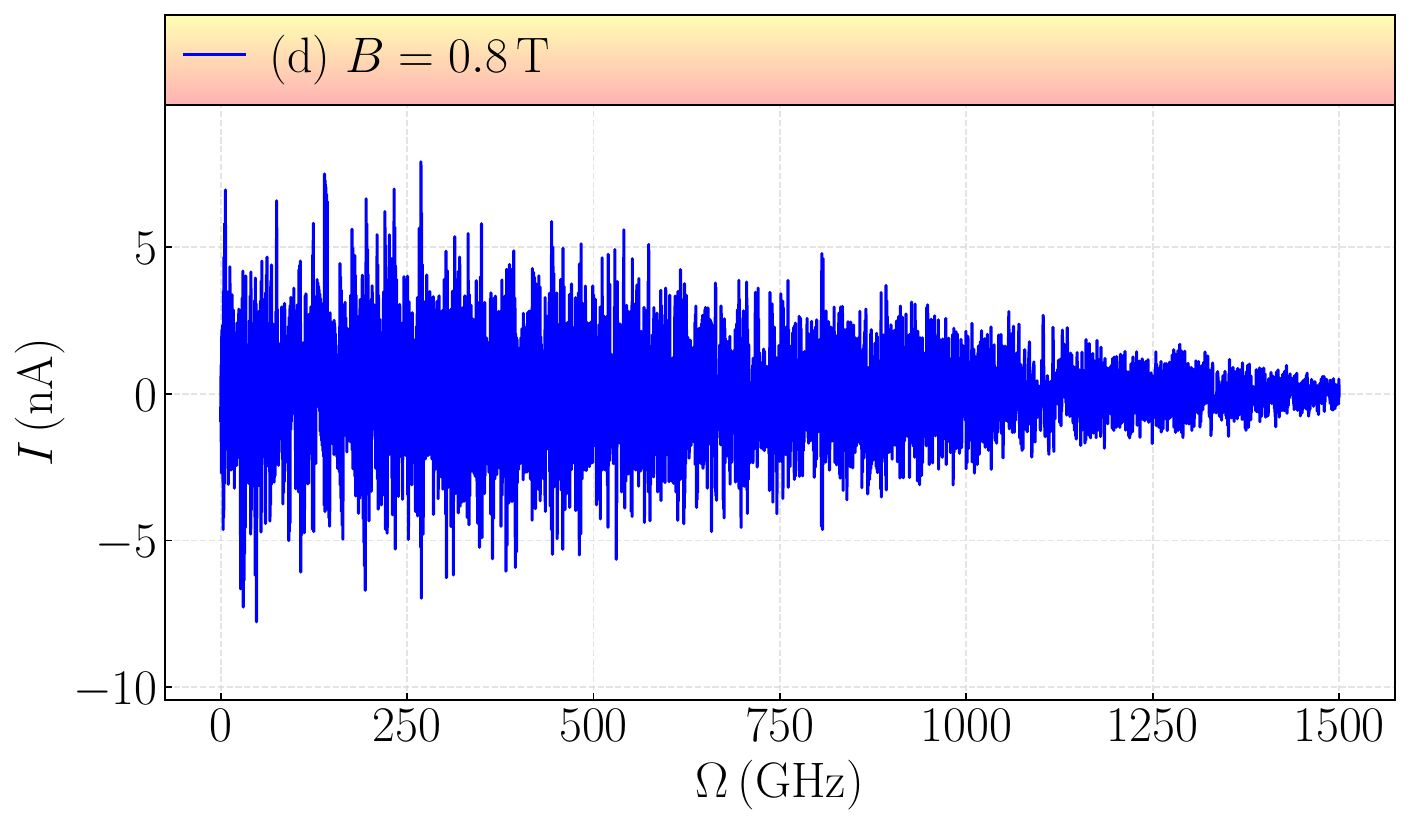}
\caption{Persistent current as a function of the rotate $\Omega$ for different values of \(B\). The parameter values used were $r_0 = 1350\,\mathrm{nm}$, $\hbar\omega_0 = 2.23\,\mathrm{meV}$.}
\label{fig:currO}
\end{center}
\end{figure*}

Starting from Eq. \eqref{eq:energy}, we investigate the magnetization of the system, a thermodynamic quantity that measures how the system (in this case, a 2D QR) responds to the application of an external magnetic field. Considering a fixed number of electrons $N_{e}$ and operating at zero temperature, the magnetization of the quantum ring is given by
\begin{equation}
    \mathcal{M}=-\left(\frac{\partial U}{\partial B}\right)_{\Omega},
\end{equation}
where
\begin{equation}
    U=\sum_{n,m}E_{n,m},
\end{equation}
is the total energy. By using Eq. \eqref{eq:energy}, the magnetization under the rotational effect is
\begin{equation}
   \mathcal{M}_{n,m}=-\frac{e\hbar}{\mu}\left[\left(n+\frac{L}{2}+\frac{1}{2}\right)\frac{\omega^*}{\omega_{\mathrm{eff}}}-\frac{1}{2}\left(m-\phi\right)\right].
\end{equation}
Making $\Omega=0$, we obtain the magnetization as described in the literature \cite{PRB.1999.8.5626}.
 \begin{figure*}[htbp]
\begin{center}
\includegraphics[width=0.48\linewidth]{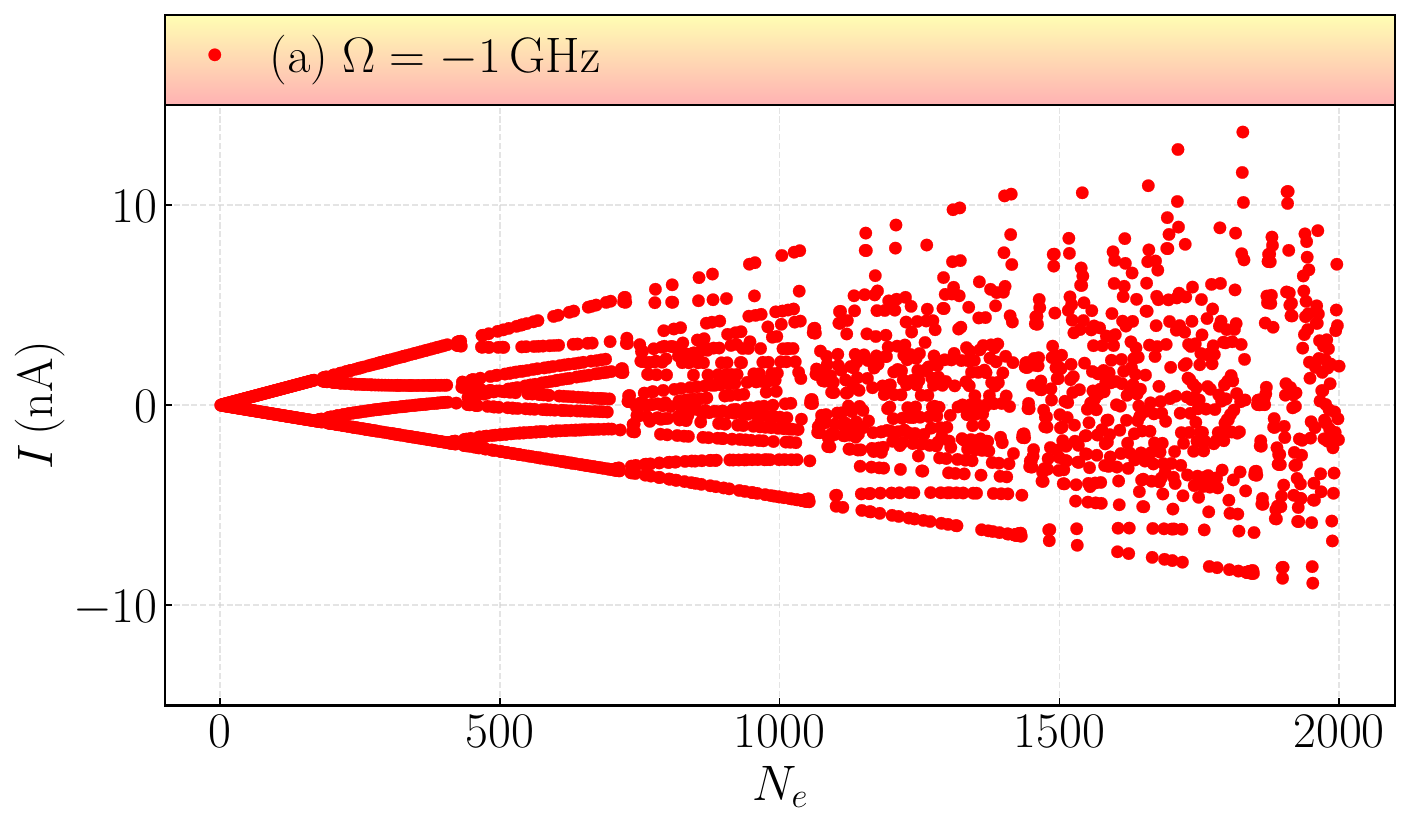}
\includegraphics[width=0.48\linewidth]{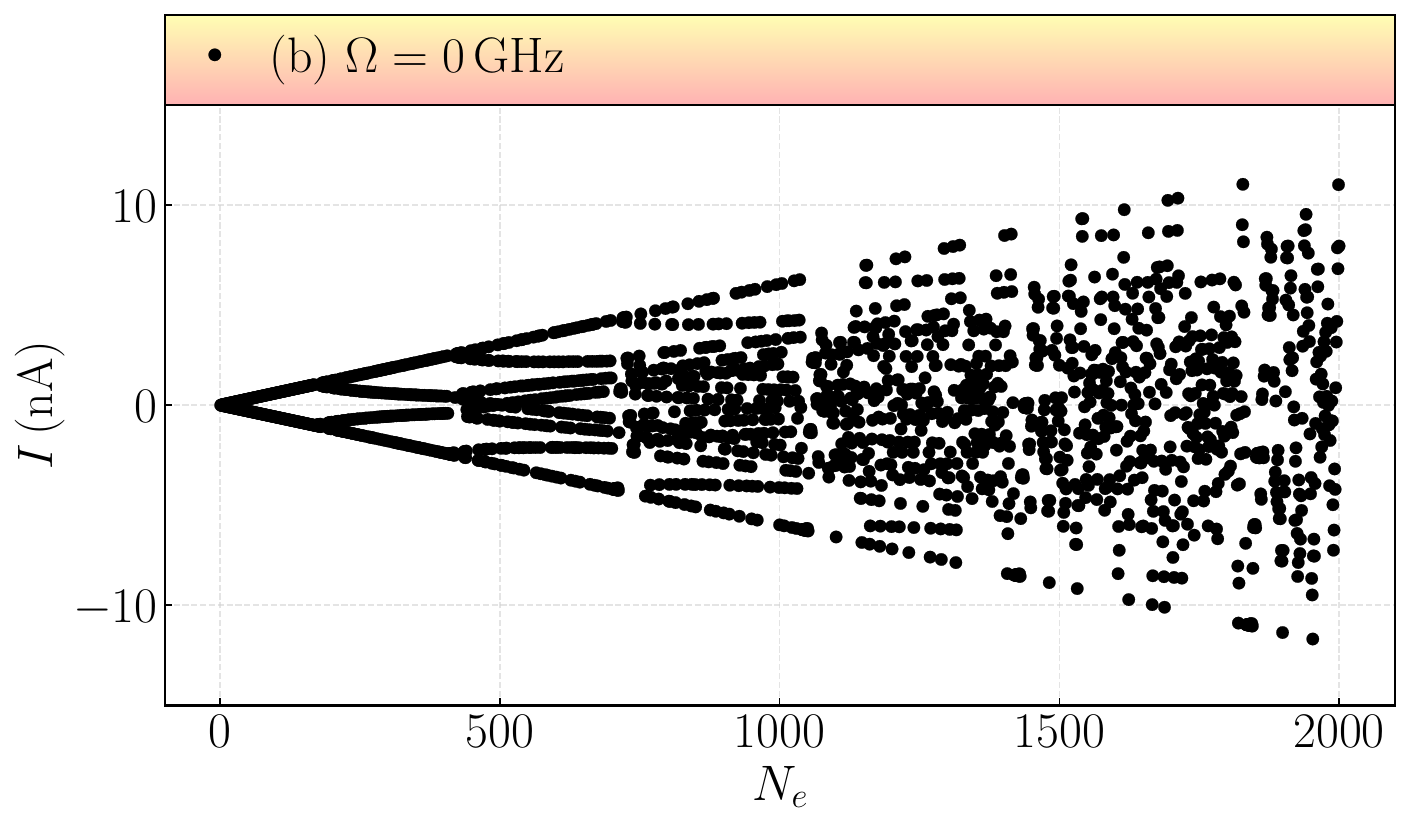}
\includegraphics[width=0.48\linewidth]{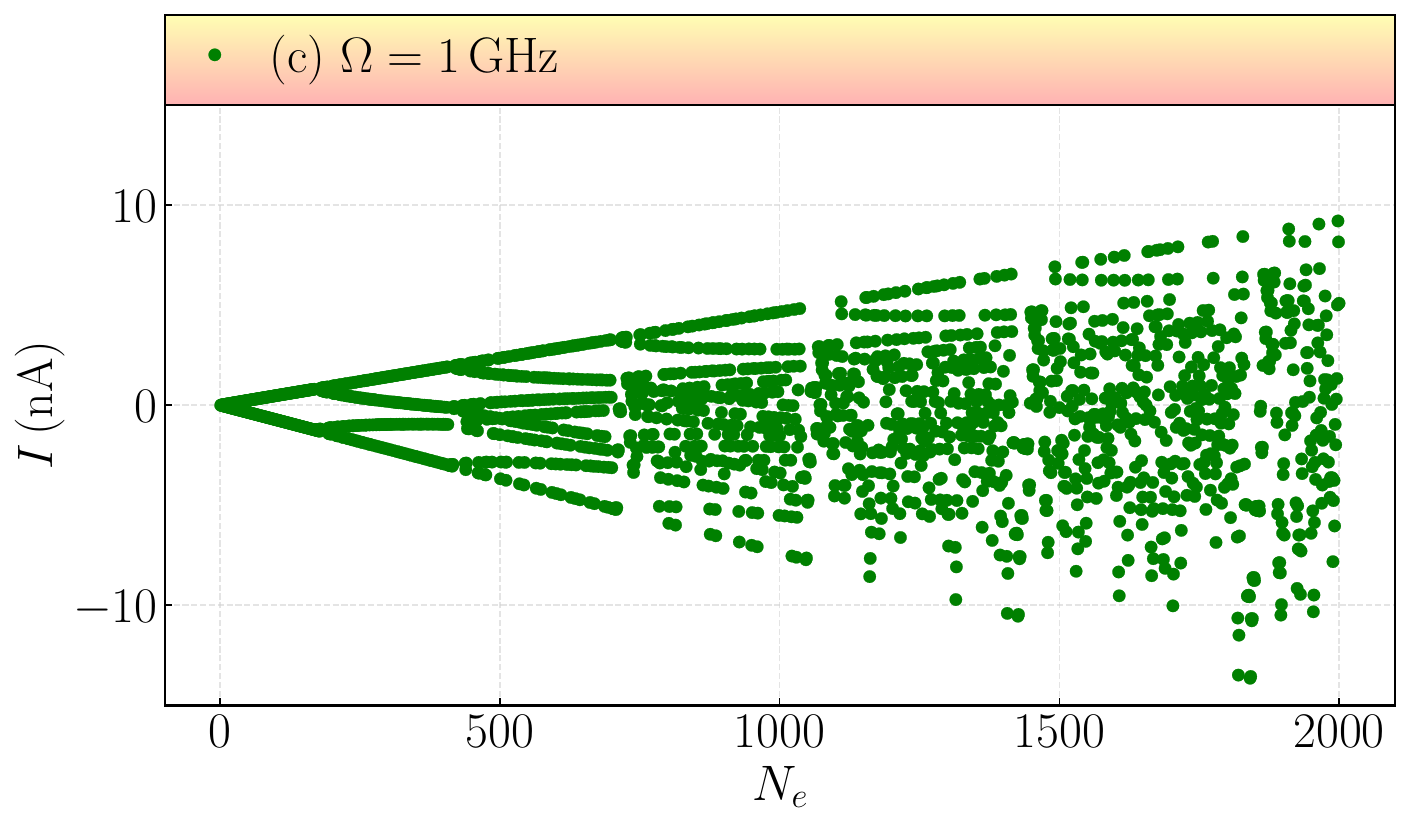}
\includegraphics[width=0.48\linewidth]{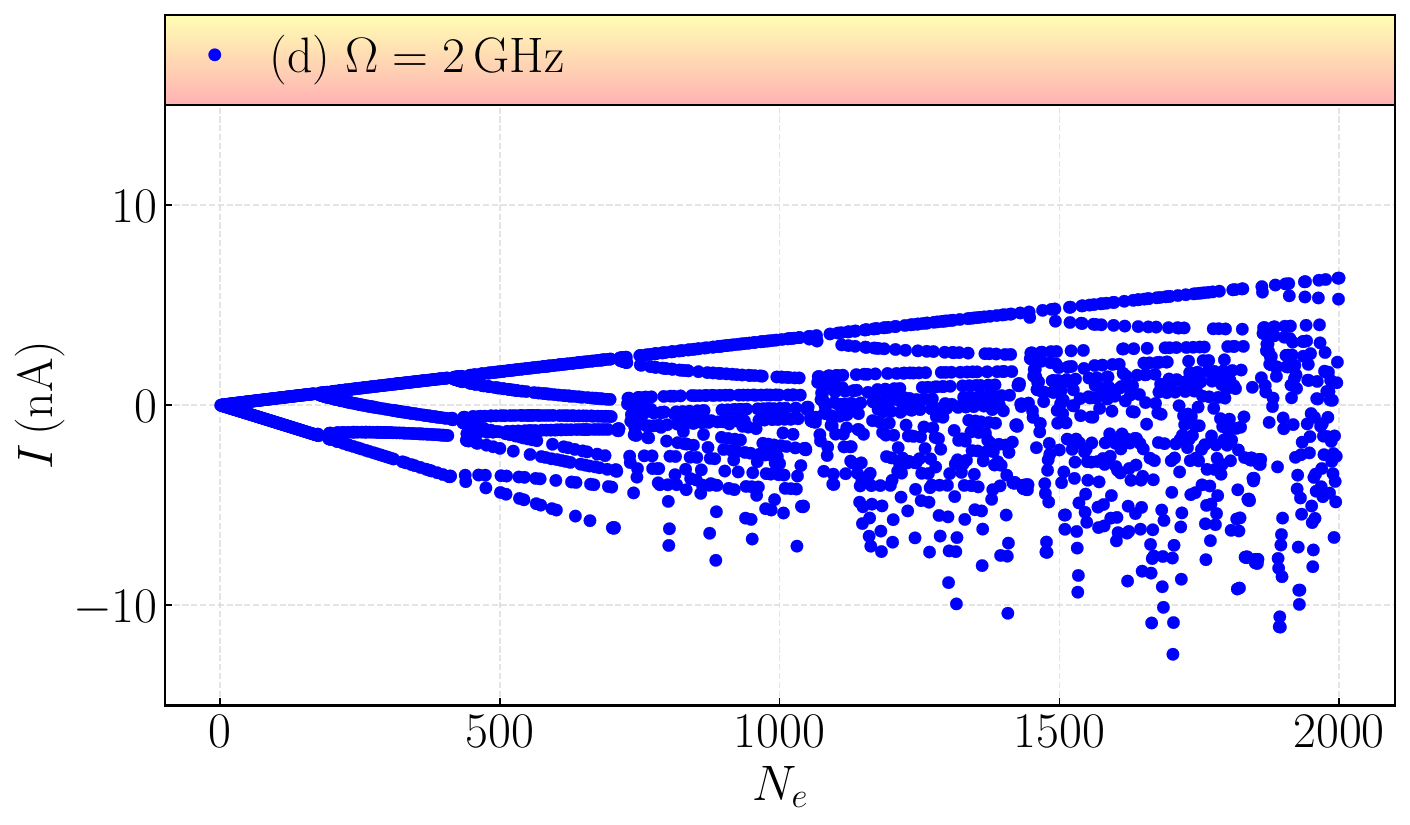}
\caption{Persistent current as a function of the number of electrons $N_{e}$ for different values of \(\Omega\). The parameter values used were $r_0 = 1350\,\mathrm{nm}$, $\hbar\omega_0 = 2.23\,\mathrm{meV}$ and $B = 1.97\times10^{-4}\,\mathrm{T}$.}
\label{fig:currN}
\end{center}
\end{figure*} 
Magnetization as a function of magnetic field has two distinct types of oscillations: AB oscillations, which predominate in the weak field regime, and de Haas-van Alphen (dHvA) oscillations, which become prominent in strong fields. AB oscillations result from the crossing of states at the Fermi energy, while dHvA oscillations result from the depopulation of subbands.

Fig.~\ref{fig:magn} shows the magnetization as a function of the field $B$ for different values of rotation $\Omega$. Rapid, low-amplitude variations characterize the AB oscillations, while the higher-amplitude oscillations correspond to the dHvA-type. In addition, AB oscillations follow the amplitude pattern of dHvA. It can be seen that, as $B$ increases, the amplitude of the AB oscillations decreases, becoming almost imperceptible at the peaks of the dHvA oscillations, while the amplitude of the dHvA oscillations increases.

As $\Omega$ increases, there is a shift in the profiles towards smaller values of $B$, which becomes clear when we analyze the maximum amplitudes of the curves. The magnetization profiles remain practically unchanged, even for very different rotation magnitudes. For example, in Fig.~\ref{fig:magn}, the four curves have been plotted for $\Omega = -300,\;0,\;150$ and $300$ GHz. One detail that can be observed is the oscillatory pattern between magnetization \ref{fig:magn} and Fermi energy \ref{fig:Efermi-omega}, evidenced by the fact that the peaks occur in the same magnetic field region; rotation, in turn, alters the dynamics of subband population and depopulation, a process that intensifies for larger rotations. As observed for the Fermi energy, the magnetization peaks reach their highest values in the range between $B = 4$ and $6~\text{T}$.

As mentioned earlier, magnetization exhibits two types of oscillations - the AB type and the dHvA type. We will analyze these oscillations in both scenarios, in the regime of weak magnetic fields, where the AB-type oscillations are more intense. Fig.~\ref{fig:magnetiw} illustrates how rotation changes the behavior of these oscillations.

For $\Omega=0$, as shown in Fig.~\ref{fig:magnetiw}(b), the standard profile described in the literature \cite{PRB.1999.8.5626} is obtained. In this case, the magnetization exhibits noise as $B$ varies, but a well-defined pattern of oscillations is still discernible, with maximum amplitudes remaining constant. As the modulus of rotation increases, this pattern begins to break down. In Fig.~\ref{fig:magnetiw}(c), for $\Omega = 150\,\mathrm{GHz}$, there is still a similarity to the profile in Fig.~\ref{fig:magnetiw}(b). However, in Figs.~\ref{fig:magnetiw}(a) ($\Omega = -300\, \mathrm{GHz}$) and \ref{fig:magnetiw}(d) ($\Omega = 300\, \mathrm{GHz}$), where the modulus of rotation is very high, the pattern is completely broken and there is a decrease in the amplitude of the oscillations, noticeable on the vertical axis of the figure.

The variation in rotation causes, for a fixed magnetic field $B$, a behavior in the magnetization profile analogous to that observed when the magnetic field is varied while keeping $\Omega$ fixed. This similarity can be verified by comparing the graphs in Figs.~\ref{fig:magn} and \ref{fig:magnO}.

In Fig.~\ref{fig:magnO}, AB and dHvA oscillations are clearly observed. In this case, we set some values of $B$ other than zero, and we note that as the magnetic field increases, the highest peaks shift to lower values of $\Omega$. Furthermore, it can be observed that for very high rotations, the AB oscillations are progressively suppressed, while the dHvA oscillations become more predominant.

A crucial difference between the graphs in Figs.~\ref{fig:magn} and \ref{fig:magnO} is in the role of the control parameters: in the graph in Fig.~\ref{fig:magn}, rotation is the control parameter and magnetization is plotted as a function of the magnetic field; in Fig.~\ref{fig:magnO}, the opposite occurs, the magnetic field is the control parameter and magnetization is analyzed as a function of rotation. We can conclude that, like the magnetic field, rotation also induces magnetization in the quantum ring, provided that $B \neq 0$.

For the scenario of magnetization in strong fields, the AB oscillations are strongly suppressed and come to be governed by the dHvA oscillations, the rotation being responsible for decreasing the magnetization range associated with this type of oscillation; in Fig.~\ref{fig:magns} this is noticeable when observing the curves moving towards progressively lower magnetization values as $\Omega$ increases, which shows how the system's response depends on the intensity of the rotation.

Let's consider that the negative sign at $\Omega=-300$ (GHz) represents clockwise rotation and the positive sign at $\Omega=300$ (GHz)  counterclockwise. It is clear that the direction of rotation produces different effects, and Fig. \ref{fig:magns} shows that the amplitude ranges for these two cases are extreme. Finally, we now explore the persistent currents and the effects of rotation on them. Starting from Eq. \eqref{eq:energy}, the generation of these currents arises from the interaction between the charge carriers and the vector potential, as a consequence of the AB effect. 

Persistent currents are numerically obtained by filling the energy levels $E_{n,m}$ up to the Fermi level and evaluating the corresponding derivatives. In short, the persistent current can be obtained through the Byers–Yang relation \cite{PRL.1961.7.46}, which describes its dependence on the magnetic flux
\begin{equation}
    I_{n, m}= -\frac{\partial E_{n, m}}{\partial \Phi}.
    \label{eq:i1}
\end{equation}
Substituting Eq. \eqref{eq:energy} into the above expression, we obtain:
\begin{equation}
    I_{n,m}=\dfrac{e\omega_{\mathrm{eff}}}{4\pi}\left(\dfrac{m-\phi}{L}-\dfrac{\omega^*}{\omega_{\mathrm{eff}}}\right),
\end{equation}
i.e., the total current is obtained by summing the contributions of all states.
\begin{equation}
    I=\sum_{n,m}I_{n,m}f_F\left(E_{n, m}\right).
\end{equation}

Fig.~\ref{fig:curr} shows the persistent current as a function of the magnetic field for different values of \(\Omega\). The rapid oscillations recorded reflect the effects of AB, whose amplitude is drastically reduced as the magnetic field increases, due to the depopulation of the subbands. When observing the vertical and horizontal scales, two behaviors can be distinguished: vertically, it can be seen that rotation attenuates the current amplitude, which decreases as the \(|\Omega|\) module increases - especially noticeable in the initial amplitude; horizontally, it can be seen that the decay of these oscillations becomes more pronounced as \(|\Omega|\) increases.

 Unlike in the case of magnetization, no oscillations of the dHvA type can be identified.

In the regime of weak magnetic fields (Fig.~\ref{fig:currw}), the amplitudes of the persistent current decrease as \(|\Omega|\) increases; the periodic character of the AB-type oscillations is evident for \(|\Omega=0|\) (Fig.~\ref{fig:currw}(b)) and gradually breaks down as \(|\Omega|\) increases. 

In Fig.~\ref{fig:currw}(c), an approximately periodic pattern is still observed, but for very large values of \(|\Omega|\) (Fig.~\ref{fig:currw}(d) and Fig.~\ref{fig:currw}(d)), the modulus of the rotation is so intense that this periodicity becomes almost imperceptible. Oscillations of the dHvA type remain absent.

For strong magnetic fields (Fig.~\ref{fig:currs}), the persistent current exhibits approximately periodic oscillations; the rotation shifts the peaks towards smaller field values as \(\Omega\) increases, as well as decreasing the magnitude of the amplitude as \(|\Omega|\) increases.

Figure \ref{fig:currO} shows the persistent current as a function of rotation for different values of the applied magnetic field: in Fig. \ref{fig:currO}(a) we have $B = 0.1\,\mathrm{T}$, in Fig. \ref{fig:currO}(b) $B = 0.4\,\mathrm{T}$, in Fig. \ref{fig:currO}(c) $B = 0.6\,\mathrm{T}$ and in Fig. \ref{fig:currO}(d) $B = 0.8\,\mathrm{T}$. It can be observed that, for very high rotation values, the amplitudes of the oscillations decrease. Analyzing the vertical scale, it can be seen that more intense magnetic fields reduce the amplitude of the current, especially noticeable in the initial value. The profile of the persistent current in Fig. \ref{fig:currO} is similar to that observed in Fig. \ref{fig:curr}, the current maintains its variation with rotation, analogous to its behavior in the magnetic field.

In Fig.~\ref{fig:currN}, we present the persistent current as a function of the number of electrons $N_{e}$ in the weak magnetic field regime and examine how rotation alters its profile. As discussed in \cite{PE.2021.132.114760}, the current exhibits oscillations in both modulus and sign, resulting from the occupation of states located to the right and left of the minimum of the subbands. 

Entry into a new subband occurs whenever two new branches appear. In Fig.~\ref{fig:currN}(b), without rotation, the same profile described by Tan-Inkson \cite{PRB.1999.8.5626} is observed. Our results, however, show that this profile undergoes significant changes even for small variations in rotation, indicating a high sensitivity of the persistent current to this parameter. As $N_{e}$ increases, it becomes difficult to distinguish the additional branching. For $\Omega = -1$ GHz (Fig.~\ref{fig:currN}(a)), the division of the branches becomes practically imperceptible at positive current values; in Fig.~\ref{fig:currN}(c), with $\Omega = 1$ GHz, this occurs at negative current values. 
Finally, at $\Omega = 2$ GHz (Fig.~\ref{fig:currN}(d)), there is an increase in the number of sub-bands occupied for negative persistent currents.

\FloatBarrier
\section{Conclusions \label{last}}

We have developed a theoretical framework to investigate the effects of rotation on key electronic properties of two-dimensional quantum rings using the Tan–Inkson confinement model. The incorporation of rotational dynamics results in significant modifications to the energy spectrum, which, in turn, affect the Fermi energy, magnetization, and persistent current. 
For the Fermi energy, we see that increasing the modulus of rotation, $\Omega$, systematically shifts the Fermi energy curves to lower values. This behavior is due to the modification of the Landau levels caused by rotation. 
Likewise, both magnetization and persistent current are also affected by the repositioning of these energy levels. In the case of magnetization, our results show that as $\Omega$ increases, the peak of the magnetization curves occurs at increasingly lower magnetic fields. This shows that rotation favors lower energy states, shifting the point of maximum magnetic response to lower field values. 
As for the persistent current, we observed a progressive decrease in the amplitude of the oscillations as $|\Omega|$ increased. This decrease is associated with the flattening of the Landau level distribution.
This approach provides a solid foundation for further numerical analysis and potential experimental comparisons in mesoscopic systems.

Importantly, our findings highlight the emergence of exotic quantum phenomena at high rotation rates (on the order of GHz). While direct experimental realization remains challenging, techniques developed for rotating Bose–Einstein condensates, such as optical lattices and tailored trapping potentials, offer promising avenues~\cite{Fetter2009}. In combination with advances in nanofabrication and artificial gauge fields, these methods present a feasible route for achieving controlled rotational dynamics in solid-state systems. The interplay between rotation and many-body interactions in quantum rings may also lead to novel states of matter, similar to those found in fractional quantum Hall systems~\cite{roncaglia2011rotating}. Thus, our theoretical predictions open new pathways for experimental validation and deepen our understanding of the rich physics of rotating quantum systems.

Before concluding, we would like to emphasize that the results presented in this article can contribute to the development of new mesoscopic devices, which exploit rotation as a control parameter, allowing for adjustments to their magnetic and current responses without the need for additional field or geometry variations.

\section*{Acknowledgments}
This work was supported by CAPES (Finance Code 001), CNPq (Grant 306308/2022-3), and FAPEMA (Grants UNIVERSAL-06395/22 and APP-12256/22).
 
\bibliographystyle{apsrev4-2}
%

\end{document}